\documentclass[aps,prc,11pt,showpacs,superscriptaddress,nofootinbib,preprintnumbers,amsmath,amssymb]{revtex4-1}
\usepackage[caption=false]{subfig}
\usepackage{setspace}
\usepackage{graphicx}
\usepackage{bm}

\begin{document}
\singlespacing

\preprint{}

\title{Chiral fluid dynamics with explicit propagation of the Polyakov loop}

\author{Christoph Herold}
\affiliation{Institut f\"ur Theoretische Physik, Goethe-Universit\"at, Max-von-Laue-Str.~1, 
60438 Frankfurt am Main, Germany}
\affiliation{Frankfurt Institute for Advanced Studies (FIAS), Ruth-Moufang-Str.~1, 60438 Frankfurt am Main, Germany}

\author{Marlene Nahrgang}
\affiliation{Frankfurt Institute for Advanced Studies (FIAS), Ruth-Moufang-Str.~1, 60438 Frankfurt am Main, Germany}
\affiliation{SUBATECH, Universit\'{e} de Nantes, EMN, IN2P3/CNRS, 4 rue Alfred Kastler, 44307 Nantes
cedex 3, France}

\author{Igor Mishustin}
\affiliation{Frankfurt Institute for Advanced Studies (FIAS), Ruth-Moufang-Str.~1, 60438 Frankfurt am Main, Germany}
\affiliation{Kurchatov Institute, National Research Center, 123182 Moscow, Russia}

\author{Marcus Bleicher}
\affiliation{Institut f\"ur Theoretische Physik, Goethe-Universit\"at, Max-von-Laue-Str.~1,
60438 Frankfurt am Main, Germany}
\affiliation{Frankfurt Institute for Advanced Studies (FIAS), Ruth-Moufang-Str.~1, 60438 Frankfurt am Main, Germany}

\date{\today}

\begin{abstract}
We present a fully dynamical model to study nonequilibrium effects in both the chiral and the deconfinement phase 
transition. The sigma field and the Polyakov loop as the corresponding order parameters are propagated by Langevin 
equations of motion. The locally thermalized background is provided by a fluid of quarks and antiquarks. 
Allowing for an exchange of energy and momentum through dissipative and stochastic processes we ensure that the total energy 
of the coupled system remains conserved. We study 
its relaxational dynamics in different quench scenarios and are able to observe critical slowing down as well as the 
enhancement of long wavelength modes at the critical point. 
During the fluid dynamical expansion of a hot plasma fireball typical nonequilibrium 
effects like supercooling and domain formation occur when the system evolves through the first order phase transition. 
\end{abstract}


\maketitle

\section{Introduction}
One of the major goals of heavy-ion physics is to gain firm knowledge about the different phases of strongly interacting 
matter and the transitions between them. 
At high temperatures, the chiral symmetry that is broken in the QCD vacuum, is expected to be restored. Furthermore, 
quarks and gluons might become the relevant degrees of freedom after the transition from hadrons to a deconfined state of 
matter.
From lattice QCD calculations we know that at vanishing baryochemical potential the chiral transition is an analytic 
crossover \cite{Aoki:2006we,Cheng:2006qk}. As recent studies show, there seems to be no direct relation between the
chiral restoration and the onset of deconfinement \cite{Aoki:2006br,Aoki:2009sc,Borsanyi:2010bp}. 
A multitude of effective models have been used to investigate the QCD phase diagram 
\cite{Scavenius:2000qd,Schaefer:2004en,Ratti:2005jh,Schaefer:2007pw,Sasaki:2007qh,Skokov:2010uh,Skokov:2010sf,Schaefer:2009ui,Sasaki:2011sd}
indicating a first order phase transition at large $\mu_B$ ending at a 
critical point (CP). In equilibrium this CP exhibits a divergence in the correlation 
length of the order parameter. The problem how to locate the CP in the $T$-$\mu_B$-plane in heavy-ion collisions was 
addressed by Stephanov, Rajagopal and Shuryak \cite{Stephanov:1998dy,Stephanov:1999zu} for equilibrated systems. They proposed to search for 
divergences in event-by-event fluctuations of quantities like transverse momentum or particle multiplicity, since the strength 
of these fluctuations is directly related to the correlation length of the sigma field, the order parameter 
for the chiral phase transition. 
However, a priori one cannot expect to reach thermodynamic equilibrium during such a heavy-ion collision, especially near the phase 
transition where relaxation times become large in comparison to the rapid dynamics of the expanding fireball. Therefore, 
several additional effects have to be taken into account. Not only is the growth of the correlation length naturally 
limited by the finite size of the system, but also critical slowing down of the dynamics in the vicinity of the critical 
point will weaken the expected signals \cite{Berdnikov:1999ph}. One may try to overcome this difficulty by looking for quantities 
that are accessible in experiment and more sensitive to the correlation length. Higher cumulants and combinations 
of them like the kurtosis of conserved quantities such as net baryon number or net charge fulfill this requirement \cite{Stephanov:2008qz,Karsch:2010ck}. 
The fast dynamical evolution of matter through a first order phase transition may exhibit interesting phenomena like supercooling followed 
by a decay through spinodal decomposition 
\cite{Csernai:1992tj,Csernai:1995zn,Scavenius:2000bb,Randrup:2010ax,Steinheimer:2012gc,Chomaz:2003dz} or nucleation 
\cite{Mishustin:1998eq}. Also an enhancement of soft pions from the decay of a disoriented chiral condensate (DCC) 
was proposed as a signal for a nonequilibrium chiral phase transition \cite{Rajagopal:1993ah,Biro:1997va,Xu:1999aq}.

Only via 
thorough theoretical understanding of the processes and effects in dynamical systems undergoing phase transitions one may 
provide realistic predictions for upcoming experiments with relativistic heavy-ion beams at RHIC 
\cite{Caines:2009yu}, CERN SPS \cite{Aduszkiewicz:2012fr}, FAIR \cite{Friman:2011zz} and NICA \cite{nica:whitepaper}. 

In \cite{Ratti:2005jh}, the thermodynamic properties of a Polyakov loop extended Nambu-Jona-Lasinio (PNJL) model were studied 
and compared with lattice QCD results. The addition of the Polyakov loop to these models improved the behavior of thermodynamic bulk 
quantities in comparison to the lattice QCD data. A similar extension to the sigma model with constituent quarks was presented in 
\cite{Schaefer:2007pw} where a coincidence of the deconfinement and chiral symmetry transition even at finite $\mu_B$ was 
found. An investigation of the phase structure of that model beyond mean-field has been performed in \cite{Skokov:2010uh} including quantum fluctuations  
within the functional renormalization group method. This improvement has lead to a quantitative shift in the position of the 
CP. Furthermore, the authors calculated net-quark number density fluctuations as well as ratios of cumulants 
as an important means to identify the location of the deconfinement and chiral phase transitions. The relevance of the fermion 
vacuum loop for such models has been investigated in \cite{Skokov:2010sf} where it was shown that this term has crucial 
influence on the phase structure and on physical observables such as net quark number fluctuations. 
In \cite{Schaefer:2009ui} an extension of this model to (2+1) flavors together with a comparison with lattice QCD data 
have been presented. An alternative model
using a dilaton field representing a scalar glueball condensate was subject to studies in \cite{Sasaki:2011sd}. 
An important step towards the understanding of nonequilibrium effects at the chiral phase 
transition was done in \cite{Sasaki:2007qh}, where it was shown that the inclusion of spinodal instabilities yields an 
enhancement of density fluctuations along the first order transition line. This is in contrast to the usual understanding 
of thermalized systems where such fluctuations are expected to diverge or grow large only at the CP.

For a better understanding of the processes during a heavy-ion collision nonequilibrium effects have to be taken into account in the 
framework of a fully dynamical model. A promising ansatz to study the QCD phase transition in such a way is given by 
a chiral fluid dynamics model that has been developed and extended over more than ten years now 
\cite{Mishustin:1998yc,Paech:2003fe,Nahrgang:2011mg,Nahrgang:2011ll,arXiv:1105.1962}. In \cite{Mishustin:1998yc}, the authors introduced a 
model coupling a relativistic ideal fluid of quarks to the linear sigma model and a scalar glueball condensate. Later 
in \cite{Paech:2003fe}
this model was used to study the $3+1$ dimensional fluid dynamic expansion of a plasma droplet coupled to the 
out-of-equilibrium evolution of the long-wavelength modes of the chiral condensate. There the initial 
fluctuations of the sigma field were propagated deterministically. A 
self-consistent derivation of the coupled dynamics of fields and fluid together with the local
thermodynamic properties of the heat bath was then presented in \cite{Nahrgang:2011mg}. This step is  
a crucial improvement as the previous models which propagated the chiral fields using a classical equation of motion  
neglected relaxational and stochastic processes. As as result, the fields would never relax to their equilibrium state for a given 
temperature but continue oscillating. In \cite{Nahrgang:2011mg}, the order parameter of chiral symmetry is propagated by a 
Langevin equation to include damping and noise in the heat bath of quarks. The quark fluid is propagated according to the 
equations of ideal hydrodynamics. 
The relaxational dynamics of the field and the treatment of the finite size of the heat bath were
investigated in \cite{Nahrgang:2011ll} and the first numerical studies of the full expanding
system were then presented in \cite{arXiv:1105.1962}. A coupling of the linear sigma model to viscous hydrodynamics has 
been studied in \cite{PeraltaRamos:2011wp}.

Dynamical models for the Polyakov loop have been proposed in \cite{Dumitru:2001,Dumitru:2002,Fraga:2007gg}. In 
\cite{Dumitru:2001}, the authors developed a model for particle production near the deconfinement phase transition due to 
oscillations in a Polyakov loop condensate. These oscillations were included on the basis of a kinetic term in an effective field theory 
for a Polyakov loop that is biquadratically coupled to a mesonic field. This idea was further used in \cite{Dumitru:2002}, 
where the authors showed how explosive behavior at the QCD phase transition might be produced by the decay of such a condensate of 
Polyakov loops. Based on work in \cite{Miller:2000pd,Meisinger:2001cq}, a Langevin equation for the deconfinement order 
parameter for pure SU(2) gauge theory was developed in \cite{Fraga:2007gg}. It was shown that  
the dissipative interaction with the medium plays a significant role in the determination of physical 
time scales. 

In the present work, we connect these ideas of a dynamical Polyakov loop with the model of chiral fluid dynamics to take into account effects 
of the deconfinement phase transition. The newly included deconfinement order parameter is treated as an effective field and 
propagated by a phenomenological Langevin equation. 
The suggested model is able to provide a dynamic description of two phase transitions 
in the background of a fluid dynamically expanding heat bath of quarks. This setup resembles the situation after the collision 
of two heavy nuclei. In \cite{Nahrgang:2011mg} the proper nonequilibrium dynamics for the sigma field and the quark fluid have been 
derived self-consistently for a model without Polyakov loop. The resulting Langevin equation for the sigma field includes two important effects: the damping of 
the chiral field due to the interaction with the heat bath and the back reaction of the heat bath on the sigma field by stochastic noise. 
For the dynamics of the Polyakov loop we follow the same idea but on a phenomenological basis as it is currently not possible to 
derive the dynamics of the Polyakov loop from a consistent field theoretical approach like it has been achieved for the sigma field. 
With the considered Polyakov loop extended linear sigma model we are able to describe characteristic phenomena at the phase boundary 
within a dynamical setup. 

This article is structured as follows. In section \ref{sec:model} we present the model of Polyakov loop extended chiral 
fluid dynamics, where the sigma field and Polyakov loop are coupled to a fluid dynamically propagated 
heat bath of quarks. After that we consider two different numerical implementations.
Section \ref{sec:equilibrationbox} presents results of relaxational dynamics in a box for several quench scenarios. 
In section \ref{sec:expandingmedium} we study the evolution of a freely expanding hot plasma undergoing a phase transition.
Finally, we present a short summary and outlook in section \ref{sec:summary}.

\section{Chiral fluid dynamics with a Polyakov loop}
\label{sec:model}

\subsection{General remarks}
Our model extends existing studies of chiral fluid dynamics 
\cite{Mishustin:1998yc,Paech:2003fe,Nahrgang:2011mg,Nahrgang:2011ll,arXiv:1105.1962} 
with the Polyakov loop to take into account effects of 
both the chiral and the deconfinement phase transition. The restoration of ${\rm SU}(N_f)_{\rm L}\times{\rm SU}(N_f)_{\rm R}$ 
chiral symmetry in the high temperature phase can be characterized by the melting of the $\langle \bar q q\rangle$ 
condensate or the sigma field, respectively. For the transition to deconfinement, the Polyakov loop $\ell$ is the 
quantity of interest. It is defined as the expectation value of the color trace of a thermal Wilson loop:
\begin{equation}
\ell=\frac{1}{N_c}\langle tr_c{\cal P}\rangle_{\beta}~,\quad \bar\ell=\frac{1}{N_c}\langle tr_c{\cal P^\dagger}\rangle_{\beta}~,
\end{equation}
where the operator ${\cal P}$ is defined as
\begin{equation}
{\cal P}=P\exp\left(i g_{\rm s} \int_0^{\beta}\mathrm d \tau A_0\right)~.
\end{equation}
Here, $P$ stands for path ordering, $A_0$ denotes the temporal component of the Euclidean color gauge field, $g_{\rm s}$ 
the strong coupling constant and $\beta=1/T$ is the inverse temperature. 
The expectation value of $\ell$ is related to the free energy $F_q(\vec x)$ of an infinitely heavy test quark 
at spatial position $\vec x$ via
\begin{equation}
\langle \ell(\vec x)\rangle=\mathrm e^{-\beta F_q(\vec x)}~.
\end{equation}
In the confined phase, this free energy diverges and therefore $\langle\ell\rangle$ vanishes whereas it takes some finite value in the 
deconfined phase. 
In the 
limit of infinitely heavy quarks, QCD is invariant under $Z(N_c)$ center symmetry transformations of the $SU(N_c)$ color gauge 
group. The Polyakov loop transforms as
\begin{equation}
\ell \rightarrow z\ell~,\quad z\in Z(N_c)~,
\end{equation}
with the consequence that the confined phase is center symmetric while this symmetry is spontaneously broken in the deconfined phase.
In the presence of dynamical quarks, the Polyakov loop is always non-zero as the free energy of a test quark does not 
diverge anymore. Nevertheless, $\langle\ell\rangle$ may still serve to characterize the transition between the two phases.

\subsection{The model}

For our studies we use the Polyakov loop extended quark meson model \cite{Schaefer:2007pw}. The Lagrangian reads
\begin{equation}
{\cal L}=\overline{q}\left[i \left(\gamma^\mu \partial_\mu-i g_{\rm s}\gamma^0 A_0\right)-g \left(\sigma +i\gamma_5 \vec\tau\cdot \vec\pi\right)\right]q + \frac{1}{2}\left(\partial_\mu\sigma\right)^2 + \frac{1}{2}\left(\partial_\mu\vec\pi\right)^2 
- U\left(\sigma,\vec\pi\right) - {\cal U}(\ell, \bar\ell)~,
\label{eq:Lagrangian}
\end{equation}
where $q=(u,d)$ is the constituent quark field, so $N_f=2$, and $\sigma$ the mesonic sigma 
field. For our simulations around the phase transition we utilize a fixed strong coupling constant of 
$\alpha_{\rm s}=g_{\rm s}^2/(4\pi)=0.3$. As we are only interested in the behavior of the order parameters, 
we neglect fluctuations of the pionic degrees of freedom and keep their values fixed at 
the vanishing expectation value $\vec\pi=\langle\vec\pi\rangle=0$ for all times. The potential for the sigma field 
reads
 \begin{equation}
U\left(\sigma\right)=\frac{\lambda^2}{4}\left(\sigma^2-\nu^2\right)^2-h_q\sigma-U_0~.
\label{eq:Usigma}
\end{equation}
The chiral symmetry of the Lagrangian (\ref{eq:Lagrangian}) is explicitly 
broken by the term $h_q$ in the potential (\ref{eq:Usigma}) taking into account the finite quark masses. The parameters in 
(\ref{eq:Usigma}) are chosen such that chiral symmetry is spontaneously broken in the vacuum, where 
$\langle\sigma\rangle=f_\pi=93$~MeV, the pion decay constant. The explicit symmetry breaking term is 
$h_q=f_\pi m_\pi^2$ with the pion mass $m_\pi=138$~MeV. These requirements lead to $\nu^2=f_\pi^2-m_\pi^2/\lambda^2$. 
Choosing $\lambda^2=19.7$ yields a realistic vacuum sigma mass $m_\sigma^2=2\lambda^2 f_\pi^2 + m_\pi^2\approx 600$~MeV. 
The constant term $U_0=m_\pi^4/(4\lambda^2)-f_\pi^2 m_\pi^2$ is chosen such that the potential energy vanishes in the 
ground state. The quark-meson coupling constant is fixed by the requirement to reproduce the constituent quark mass in 
vacuum, $g=m_q/f_\pi=3.3$.

The temperature dependent Polyakov loop potential is chosen in a polynomial form \cite{Schaefer:2007pw,Ratti:2005jh,Pisarski:2000eq}.
\begin{equation}
\frac{{\cal U}}{T^4}\left(\ell, \bar\ell\right)= -\frac{b_2(T)}{4}\left(\left|\ell\right|^2+\left|\bar\ell\right|^2\right)-\frac{b_3}{6}\left(\ell^3+\bar\ell^3\right) + \frac{b_4}{16}\left(\left|\ell\right|^2+\left|\bar\ell\right|^2\right)^2~.
\label{eq:Uloop}
\end{equation}
At non-vanishing $\mu_B$, this parametrization yields an unphysical behavior in the Polyakov loop susceptibilities which 
become negative in a broad temperature range \cite{Sasaki:2006ww}. One can cure this problem by augmenting this effective
potential with a logarithmic term to account for the Haar measure in the group integral 
\cite{Sasaki:2006ww,Ratti:2007jf,Roessner:2006xn}. As we will restrict our numerical studies to the case of zero 
baryochemical potential, we can ignore this issue for the present investigation.

The coefficients in (\ref{eq:Uloop}) are fixed to reproduce thermodynamic results from lattice QCD simulations in the pure gauge sector. 
We use parametrizations proposed in \cite{Ratti:2005jh,Ratti:2004ra,Heinzl:2005xv,deForcrand:2001nd,Wozar:2006fi}
\begin{equation}
 b_2(T) = a_0 + a_1\left(\frac{T_0}{T} \right) + a_2\left(\frac{T_0}{T} \right)^2 + a_3\left(\frac{T_0}{T} \right)^3
\end{equation}
with $a_0=6.75$, $a_1=-1.95$, $a_2=2.625$, $a_3=-7.44$ and two temperature independent coefficients $b_3=0.75$ and 
$b_4=7.5$. The potential (\ref{eq:Uloop}) has a first-order phase transition at a critical temperature of $T_0=270$~MeV. 
However, in the presence of dynamical quarks this transition temperature gets lower due to the increased number of 
degrees of freedom and reduced dynamical scale. The proper $N_f$- and $\mu_B$-dependence 
of $T_0$ has been investigated in \cite{Herbst:2010rf}. For two flavors and vanishing baryochemical potential, $T_0$ reduces 
to a value of $208$~MeV.

\begin{figure}[htb!]
\centering
   \subfloat[\label{fig:fopot}]{
   \centering
   \includegraphics[scale=0.4, angle=270]{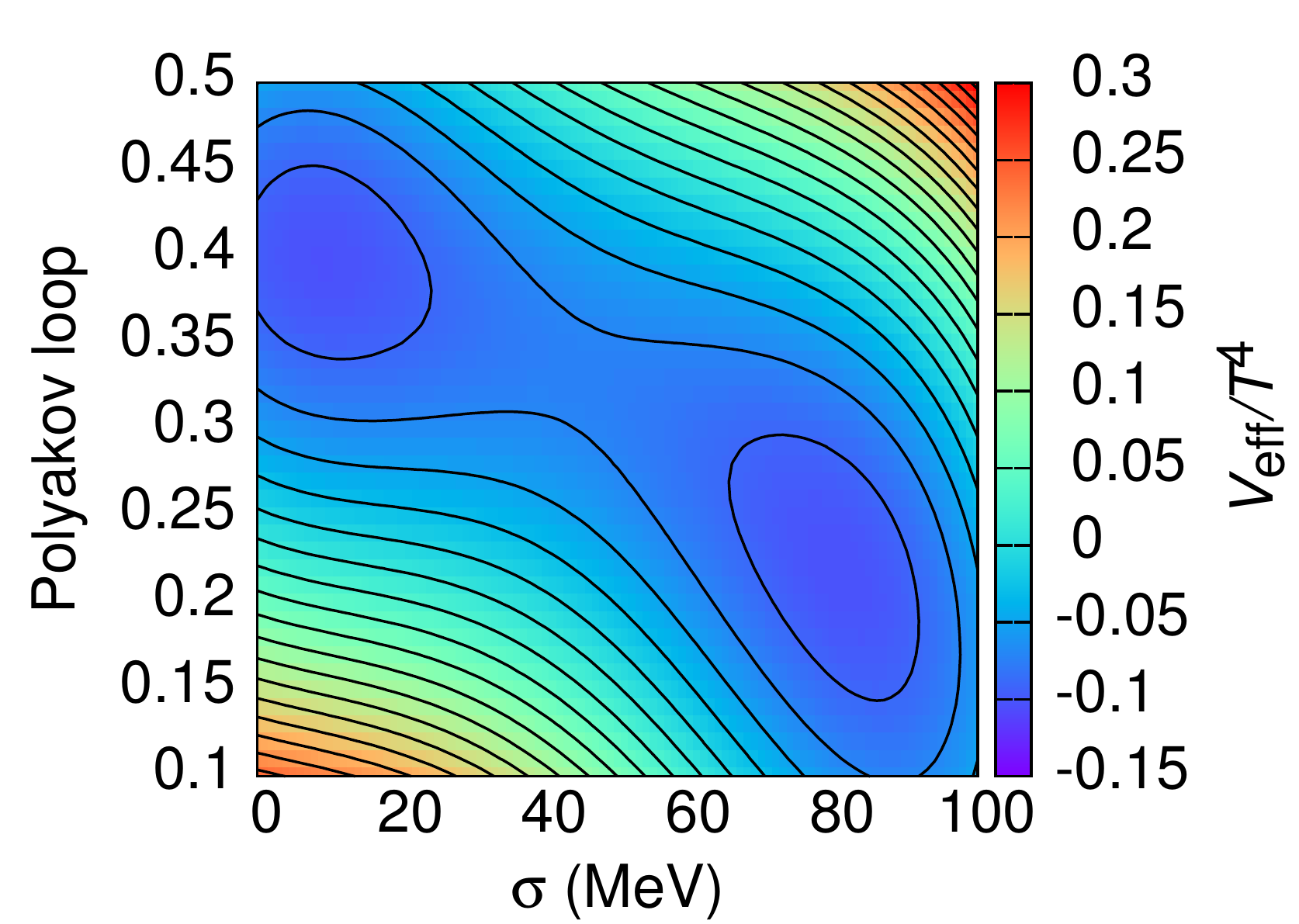}
   }
 \qquad
   \subfloat[\label{fig:cppot}]{
   \centering
   \includegraphics[scale=0.4, angle=270]{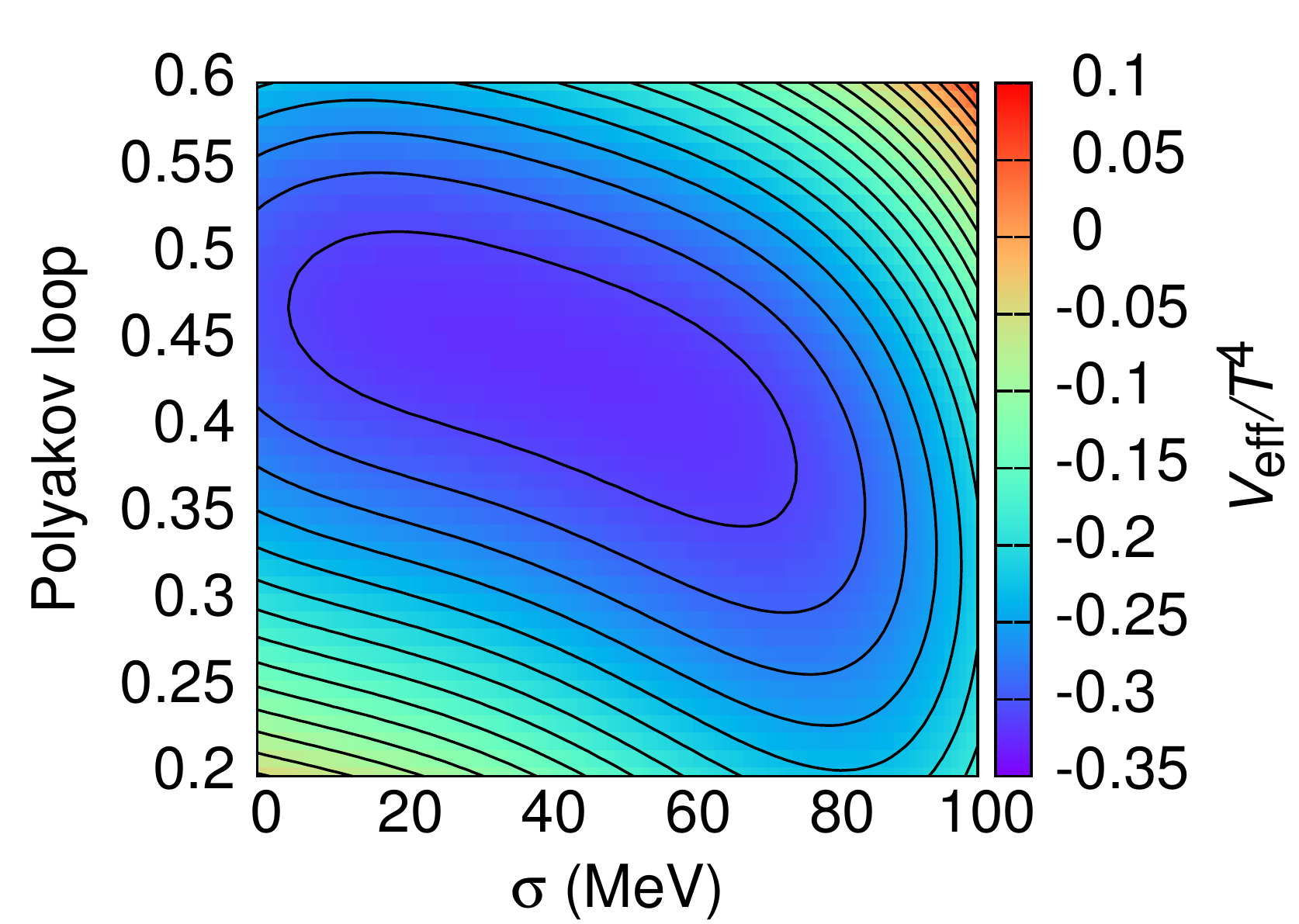}
   }
 \caption{\ref{fig:fopot} Effective potential for $g=4.7$, corresponding to a first order phase transition at $T_c=172.9$~MeV.
 \ref{fig:cppot} Effective potential for $g=3.52$, corresponding to a CP scenario at $T_c=180.5$~MeV.}
\label{fig:potential}
\end{figure}

The effective thermodynamic potential that we need for describing the dynamics of $\sigma$ and $\ell$ and for the local 
equilibrium properties of the quarks is given by
\begin{equation}
 V_{\rm eff}=-\frac{T}{V}\ln {\cal Z}=U+{\cal U}+\Omega_{q\bar q}~.
\end{equation}
The partition function ${\cal Z}$ of our system can be written as a path integral over the quarks, antiquarks, mesons and 
the temporal component of the color gauge field. Integrating out the quark degrees of freedom, which will constitute the 
heat bath, we obtain the grand canonical potential $\Omega_{q\bar q}$. At $\mu_B=0$, $\ell= \bar \ell$ and in the 
mean-field approximation it is \cite{Schaefer:2007pw}
\begin{equation}
 \Omega_{q\bar q}=-4 N_f T\int\frac{\mathrm d^3 p}{(2\pi)^3} \ln\left[1+3\ell\mathrm e^{-\beta E}+3\ell\mathrm e^{-2\beta E}+\mathrm e^{-3\beta E}  \right]~.
\label{eq:grandcanonical}
\end{equation}
Here $E=\sqrt{p^2+g^2\sigma^2}$, the energy of the quarks, their mass being generated dynamically by the sigma field. 
The integrand contains contributions from one-, two- and three-quark states, the first two being proportional to $\ell$. 
This means that for vanishing value of the Polyakov loop only three-quark states contribute, while the amount of one- and 
two-quarks gets larger with growing $\ell$. This is called ``statistical confinement'' \cite{Morita:2011jv}. In (\ref{eq:grandcanonical}), we 
omit the zero-temperature contribution to $\Omega_{q\bar q}$
which can partly be renormalized into the parameters $\lambda^2$ and $\nu^2$, leaving a logarithmic term 
depending on the renormalization scale and the effective quark mass. This term may have crucial influence on the phase 
structure of the model \cite{Skokov:2010sf,Mocsy:2004ab}. However, as the mean-field approximation provides us with the desired phase 
transition already \cite{Scavenius:2000qd}, we neglect this term and its effects in the following. In order to simplify the 
calculations and for a first qualitative study we follow the same strategy as in \cite{Paech:2003fe,Nahrgang:2011mg,Nahrgang:2011ll}.
Varying the coupling strength $g$ one can tune the characteristic shape of the effective potential $V_{\rm eff}$ at $\mu_B=0$
and by that the type of transition: 
For $g=4.7$ we see two degenerate minima $(\sigma=9~\rm{MeV},\ell=0.40)$ and $(\sigma=81~\rm{MeV},\ell=0.22)$ at the 
transition temperature of $T_c=172.9$~MeV, see fig.\ \ref{fig:fopot}. While for $g=3.52$ we have only one single minimum 
$(\sigma=49~\rm{MeV},\ell=0.43)$ at $T_c=180.5$~MeV, where 
the potential is very broad and flat as shown in fig.\ \ref{fig:cppot}. This resembles a CP. Note that in principle one 
has to choose $g$ such that the product $g\sigma$ in vacuum reproduces the constituent quark mass, which leads to a value 
of $g=3.3$. 

\subsection{The equations of motion}
\label{sec:eqmotion}

In \cite{Nahrgang:2011mg} we have derived the coupled dynamics of the sigma field and the quark fluid self-consistently 
with the two particle irreducible (2PI) effective action for an analogous model without Polyakov loop. 
The description of nonequilibrium processes can be achieved via Langevin equations. This has been extensively done in a
quantum field theoretical framework for $\phi^4$ theory \cite{Morikawa:1986rp,Gleiser:1993ea,Boyanovsky:1996xx,Greiner:1996dx}, 
gauge theories \cite{Bodeker:1995pp,Son:1997qj} and $\mathcal O(N)$ chiral models \cite{Rischke:1998qy}. Here, a splitting 
between the long- and short-wavelength modes of the sigma field was assumed and the relaxational dynamics of the soft 
modes in the heat bath of hard modes was derived within the influence functional formalism. 
Utilizing a chiral model with constituent quarks, we assumed a different splitting. Taking the quarks 
as the environmental degrees of freedom and the sigma as the relevant degrees one can calculate the 2PI effective action 
by integrating over the Keldysh contour. Out of that we were able to derive a Langevin equation of motion containing 
friction and noise\footnote{In this form the equation of motion is not Lorentz invariant. 
The problem could be cured by replacing $\eta_{\sigma}\partial_t \sigma\rightarrow \eta_{\sigma}u^{\mu}\partial_{\mu} \sigma$.}

\begin{equation}
 \partial_{\mu}\partial^{\mu}\sigma+\eta_{\sigma}(T)\partial_t \sigma+\frac{\partial V_{\rm eff}}{\partial\sigma}=\xi_{\sigma}~ .
\label{eq:eomsigma}
\end{equation}

The damping coefficient $\eta_{\sigma}$ is temperature dependent and responsible for the transfer of energy-momentum to 
the heat bath. 
It arises from the $\sigma\leftrightarrow q \bar q$ process and is non-vanishing wherever this decay is kinematically 
possible.
In \cite{Nahrgang:2011mg} we derived its explicit form in the $\left|\vec k\right|=0$ limit which is 
sufficient as we are interested mainly in the fluctuations of the soft modes. For $m_{\sigma}>2m_q$ it is given by
\begin{equation}
 \eta_{\sigma}=\frac{12 g^2}{\pi}\left[1-2n_{\rm F}\left(\frac{m_\sigma}{2}\right)\right]\frac{\left(\frac{m_\sigma^2}{4}-m_q^2\right)^{\frac{3}{2}}}{m_\sigma^2}~, 
\label{eq:dampcoeff}
\end{equation}
with $m_\sigma$ being the pole mass of the sigma field. While the screening mass, defined as the curvature of the effective 
potential at the equilibrium value, vanishes at the CP, this is not in general true for the pole mass. 
However, we use 
\begin{equation}
 m_\sigma^2=\left. \frac{\partial^2 V_{\rm eff}}{\partial\sigma^2} \right|_{\sigma=\sigma_{eq}, \,\ell=\ell_{eq}}
\end{equation}
as an approximation to the proper pole mass for equation (\ref{eq:dampcoeff}). This ansatz is justified as the pole mass 
has a minimum at the transition temperature for a crossover scenario just like the screening mass. A consistent calculation 
is possible by considering the dispersion relations for different types of excitations. 

\begin{figure}[htb!]
\centering
\includegraphics[scale = 0.55]{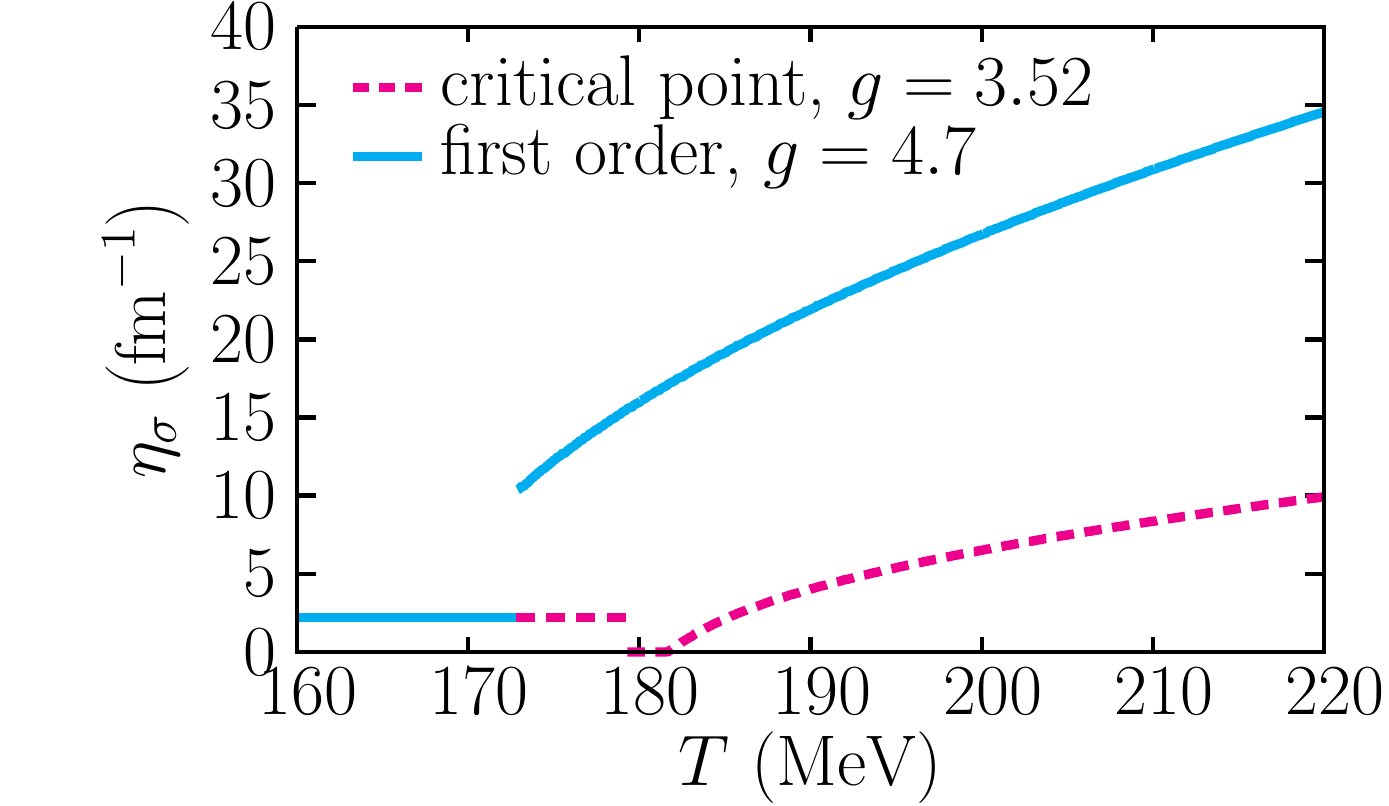}
\caption{Damping coefficient for the sigma field as a function of temperature, for a CP and a first order transition scenario.}
\label{fig:dampingcoeff}
\end{figure}

For the damping below the phase transition we choose $\eta_{\sigma}=2.2/{\mathrm fm}$ \cite{Biro:1997va} to account for the 
$\sigma\leftrightarrow \pi \pi$ reaction whenever it is kinematically allowed. Fig.\ \ref{fig:dampingcoeff} shows  
$\eta_{\sigma}$ as a function of temperature for both transition scenarios. Due to its perturbative derivation, 
the damping coefficient is rather high, especially for the first order transition. The only region where it vanishes is 
around $T_c$ for the CP, where the sigma becomes light enough so that $m_{\sigma}<2m_q,\ 2m_{\pi}$.
The stochastic noise field in the equation of motion (\ref{eq:eomsigma}) has a 
vanishing expectation value $\langle\xi_{\sigma}\rangle=0$. To complete this part we derive a dissipation-fluctuation-relation that connects 
the damping coefficient to the correlator of the stochastic noise field and ensures the relaxation to the proper 
equilibrium state.

\begin{equation}
\label{eq:correlationnoise}
 \langle\xi_{\sigma}(t,\vec x)\xi_{\sigma}(t',\vec x')\rangle=\frac{1}{V}\delta(t-t')\delta(\vec x-\vec x')m_\sigma\eta_{\sigma}\coth\left(\frac{m_\sigma}{2T}\right)~.
\end{equation} 

The temperature, the mass of the sigma field, the damping and the noise field in (\ref{eq:correlationnoise}) are local 
quantities which may vary between different positions on the spatial grid. 

The derivation of the equation of motion for the Polyakov loop $\ell$ is more problematic because $\ell$ is defined 
in the Euclidean space and it is not completely clear how to propagate it in real time.
In \cite{Dumitru:2001} a kinetic term for the Polyakov loop has been included in the Lagrangian:

\begin{equation}
 \mathcal{L}\rightarrow\mathcal{L}+ \frac{N_c}{g_{\rm s}^2}\left|\partial_{\mu}\ell\right|^2 T^2~.
\end{equation}
It was proposed that production of pions near the phase transition is driven by oscillations of the Polyakov loop. 
Only in the region around $T_c$ the Polyakov loop is light, allowing large fluctuations and thus particle production.
As the Polyakov loop operator is a phase in color space and therefore $\ell$ a pure number, dimensions in an 
effective Lagrangian can only be made up by powers of the temperature $T$. However, the temperature in our model is 
space and time dependent, so the Euler-Lagrange equation for $\ell$ would contain derivative terms of $T$ that might 
lead to unphysical behavior. 

For our present model we follow a different strategy formulated in \cite{Landau:1983}. If the order parameter is out of equilibrium, then 
relaxational processes towards its thermodynamic equilibrium value will occur. The velocity of relaxation is then 
proportional to the derivative of the effective potential with respect to the order parameter. Analogous to the sigma field 
we include fluctuations of the Polyakov loop with a stochastic noise term for which $\langle\xi_{\ell}\rangle=0$: 

\begin{equation}
 \eta_{\ell}\partial_t \ell T^2+\frac{\partial V_{\rm eff}}{\partial\ell}=\xi_{\ell}
\label{eq:eomell}~,
\end{equation}
\begin{equation}
 \langle\xi_{\ell}(t,\vec x)\xi_{\ell}(t',\vec x')\rangle T^2=\frac{1}{V}\delta(t-t')\delta(\vec x-\vec x')2\eta_{\ell} T~.
\label{eq:dissfluctpolyakov}
\end{equation} 
Equation (\ref{eq:dissfluctpolyakov}) assumes Gaussian and Markovian noise as it was used in \cite{Fraga:2007gg}.
In contrast to the sigma field it is not clear how to derive a damping for the Polyakov loop in a field theoretical manner 
as the fermionic part of the Lagrangian contains interaction with the $A_0$ color gauge field but not directly 
with $\ell$. Therefore we chose a `reasonable' value of $\eta_{\ell}= 5/fm$, although the results are not sensitive to it. 
We will further comment on this later.

\subsection{Propagation of the quark fluid}

The propagation of the quark fluid is governed by the equations of ideal relativistic fluid dynamics
\begin{equation}
 \partial_{\mu}\left(T^{\mu\nu}_q+T^{\mu\nu}_{\sigma}+T^{\mu\nu}_{\ell}\right)=0~.
\label{eq:fluiddynamics}
\end{equation}
We require energy-momentum conservation for the system as a whole, its energy-momentum tensor being split into quark, sigma 
and Polyakov loop parts. In \cite{Nahrgang:2011mg} we were able to derive self-consistently an approximate expression for 
the quark and sigma contribution. For the quarks we obtained the energy-momentum tensor of an ideal fluid, 
the contribution of the sigma field can be expressed as
\begin{equation}
 \partial_{\mu}T^{\mu\nu}_{\sigma}=\left(-\frac{\partial \Omega_{q\bar q}}{\partial \sigma}-\eta_{\sigma}\partial_t \sigma\right)\partial^{\nu}\sigma~.
\label{eq:sigmasource}
\end{equation}
Assuming the same structure and effects for the Polyakov loop contribution, we end up with
\begin{equation}
 \partial_{\mu}T^{\mu\nu}_{\ell}=\left(-\frac{\partial \Omega_{q\bar q}}{\partial \ell}-\eta_{\ell}\partial_t \ell T^2\right)\partial^{\nu}\ell~.
\label{eq:loopsource}
\end{equation}
As explained in \cite{Nahrgang:2011mg}, these terms cannot account for the average energy transfer from the heat 
bath to the fields caused by the stochastic noise fields $\xi_{\sigma}$ and $\xi_{\ell}$.
The equation of state $p=p(e)$ is obtained and tabulated from the thermodynamic relations
\begin{eqnarray}
\label{eq:pressure}
 p(\sigma, \ell, T)&=&-\Omega_{q\bar q}(\sigma, \ell, T)~,\\
\label{eq:energydensity}
 e(\sigma, \ell, T)&=&T\frac{\partial p(\sigma, \ell, T)}{\partial T}-p(\sigma, \ell, T)~.
\end{eqnarray}
A simple relation for $p(e)$ cannot be obtained as $\sigma$ and $\ell$ are propagated explicitly and can be out of equilibrium 
and the pressure as well as the energy density of the quarks depend explicitly on the local values of these fields.

\section{Equilibration in a box}
\label{sec:equilibrationbox}

In this section we study the relaxational behavior of the order parameters in a box of finite size after several 
temperature quenches. The aim is to give estimates and compare relaxation times near and far away from the transition 
point for a first order and a CP scenario. Furthermore we investigate fluctuations during and after the 
equilibration process.

\subsection{Numerical implementation}
We solve the equations of motion for the fields and the quark fluid on a fixed spatial cube of size $L^3$ where $L=N\Delta x$ 
with $N$ equal to the number of cells in each direction and the grid spacing $\Delta x=0.2$~fm. 
The fluid dynamic equations (\ref{eq:fluiddynamics}) are solved using the full 
(3+1)d SHarp And Smooth Transport Algorithm (SHASTA) ideal fluid dynamic code \cite{Rischke:1995ir,Rischke:1995mt}. To ensure 
numerical stability we use a time step of $\Delta t = 0.4\cdot\Delta x = 0.08$~fm. 

Rewriting the equation of energy and momentum conservation for the coupled system with a source term $S^{\nu}$ for the 
fluid we obtain
\begin{equation}
 \partial_{\mu}T^{\mu\nu}_q=-\partial_{\mu}\left(T^{\mu\nu}_{\sigma}+T^{\mu\nu}_{\ell}\right)=S^{\nu}~.
\label{eq:fluiddynamicssource}
\end{equation}
After performing the fluid dynamical step for vanishing $S^{\nu}$ in the standard fashion, we subtract the sources from the 
energy and momentum density in the global rest frame of the fluid. This is especially interesting for an expanding medium 
and has been successfully implemented in previous studies of chiral fluid dynamics simulations without a Polyakov loop 
\cite{Nahrgang:2011ll,arXiv:1105.1962}. As will be shown later, this enables us to conserve the total 
energy for our Polyakov loop extended model, see section \ref{sec:energy}.

To implement the source term in the numerical simulation we apply the same strategy as in \cite{arXiv:1105.1962}. Equations 
(\ref{eq:sigmasource}) and (\ref{eq:loopsource}) provide us with the energy density dissipated into the heat bath.
\begin{equation}
 \Delta e_{\rm{diss}}=\left[\left(\frac{\partial \Omega_{q\bar q}}{\partial \sigma}+\eta_{\sigma}\partial_t \sigma\right)\partial_t\sigma
+\left(\frac{\partial \Omega_{q\bar q}}{\partial \ell}+\eta_{\ell}\partial_t \ell T^2\right)\partial_t\ell \right]\Delta t~.
\end{equation}
The energy transfer due to stochastic fluctuations $\Delta e_{\rm fluc}$ can be calculated by comparing this term to 
the numerically obtained energy difference in the sigma and Polyakov loop fields before and after each time step. 
The energy density of the sigma field is given by the sum of a potential, kinetic and fluctuation term while 
for the Polyakov loop it is associated only with the potential energy due to the lack of a kinetic term in the Lagrangian. 
\begin{eqnarray}
 e_{\sigma}&=&U(\sigma)+\frac{1}{2}\left(\frac{\partial \sigma}{\partial t}\right)^2+\frac{1}{2}\left(\vec\nabla\sigma\right)^2\label{eq:energysigmafield}~,\\
 e_{\ell}&=&{\cal U}(\ell)~.
\label{eq:energypolyakovfield}
\end{eqnarray}
Altogether this gives us the zero component of the source term 
\begin{equation}
 S^0=\frac{1}{\Delta t}\left(\Delta e_{\rm diss}+\Delta e_{\rm fluc}\right)~.
\end{equation}
The spatial components accounting for momentum transfer are calculated in an analogous fashion. The momentum transfer due 
to dissipative processes is given by 
\begin{equation}
 \Delta \vec M_{\rm diss}=\left[\left(\frac{\partial \Omega_{q\bar q}}{\partial \sigma}+\eta_{\sigma}\partial_t \sigma\right)\vec\nabla\sigma
+\left(\frac{\partial \Omega_{q\bar q}}{\partial \ell}+\eta_{\ell}\partial_t \ell T^2\right)\vec\nabla\ell \right]\Delta t~,
\end{equation}
and $\Delta\vec M_{\rm fluc}$ is obtained by comparing this to the change in the momentum density of the fields
which is vanishing for the Polyakov loop. 
\begin{eqnarray}
 \vec M_{\sigma}&=&\partial_t \sigma \vec\nabla \sigma~, \\
 \vec M_{\ell}&=&0~.
\label{eq:momentumfields}
\end{eqnarray}
This gives us the spatial part of the source term
\begin{equation}
 \vec S=\frac{1}{\Delta t}\left(\Delta \vec M_{\rm diss}+\Delta \vec M_{\rm fluc}\right)~.
\end{equation}

The Langevin equations of motion for the fields $\sigma$ and $\ell$ are solved using a staggered leap-frog algorithm, cf. 
\cite{CassolSeewald:2007ru}. For this algorithm the time step is chosen smaller than for the propagation of the fluid. We use a four times smaller 
$\Delta t$ of $0.1\cdot\Delta x = 0.02$~fm. I.~e., four consecutive time steps of calculation for the fields are performed with the 
same fluid background, followed by one step for the fluid propagation. After that, the local temperatures $T(\vec x)$ 
are adjusted in each cell via a root finder of 
\begin{equation}
 e_{\rm fluid}(\vec x)-e\{\sigma(\vec x), \ell(\vec x), T(\vec x)\} = 0~.
\label{eq:temproot}
\end{equation}
These local temperatures then enter the local potentials (\ref{eq:Uloop}), (\ref{eq:grandcanonical}) and 
consequently the equations of motion (\ref{eq:eomsigma}) and (\ref{eq:eomell}) which are then used in the next 
time step for the propagation of the order parameter fields. 

\subsection{Results}
We are interested in the investigation of relaxational processes of the coupled system of fields and fluid after a temperature quench. 
Using a cubic $64^3$ grid with periodic boundary conditions we set the temperature $T_{\rm ini}$ uniformly to a value above $T_c$ and 
then initialize the sigma and Polyakov loop field at their equilibrium values including thermal fluctuations with a  
variance given by
\begin{eqnarray}
 \langle\delta\sigma^2\rangle&=&\frac{T}{V}\frac{1}{m_{\sigma}^2}~,\\
 \langle\delta\ell^2\rangle&=&\frac{T}{V}\frac{1}{m_{\ell}^2}~.
\label{eq:fluctuations}
\end{eqnarray}
Here we have defined the mass of the Polyakov loop $m_{\ell}$ analogously to the sigma mass as
\begin{equation}
m_{\ell}^2=\left. \frac{1}{T^2}\frac{\partial^2 V_{\rm eff}}{\partial\ell^2}\right|_{\sigma=\sigma_{eq}, \,\ell=\ell_{eq}}~.
\end{equation}
We then quench the temperature to various values $T<T_c$ and initialize the quark heat bath by calculating its 
energy density and pressure out of the given quantities $T$, $\sigma$, $\ell$ via equations (\ref{eq:pressure}) and
(\ref{eq:energydensity}). After that we let the system evolve 
according to equations (\ref{eq:eomsigma}), (\ref{eq:eomell}) and (\ref{eq:fluiddynamicssource}). Fields and fluid now influence 
each other in the following way: The amount of energy that the fields lose through damping gets transferred to the fluid which 
causes an adjustment of the temperature on account of equation (\ref{eq:temproot}). This new temperature then reshapes the 
thermodynamic potential that influences the dynamics of the fields. In this kind of box calculations pressure gradients in the 
fluid are small, so we expect the dynamics to be governed by the fields.

We are now interested in the relaxational behavior of the sigma field and the Polyakov loop comparing volume averages 
over all cells of the box as they evolve in time for 
different quench temperatures and two transition scenarios. The volume averages in a single event are defined as
\begin{equation}
 \langle\sigma\rangle=\frac{1}{N^3}\sum_{ijk} \sigma_{ijk}~, \ \langle\ell\rangle=\frac{1}{N^3}\sum_{ijk} \ell_{ijk}~,
\end{equation}
where $\sigma_{ijk}$ ($\ell_{ijk}$) is the instantaneous value of the sigma field (Polyakov loop) in a cell with 
coordinates i, j, k.
These values are furthermore averaged over $N_e$ events with different noise configurations
\begin{equation}
 \overline{\langle\sigma\rangle}=\frac{1}{N_e}\sum_{n=1}^{N_e} \langle\sigma\rangle_{n}~, \ \overline{\langle\ell\rangle}=\frac{1}{N_e}\sum_{n=1}^{N_e} \langle\ell\rangle_{n}~.
\end{equation}

\begin{figure}[htbp]
\centering
  \subfloat[\label{fig:forelax}]{
  \includegraphics[scale=0.5]{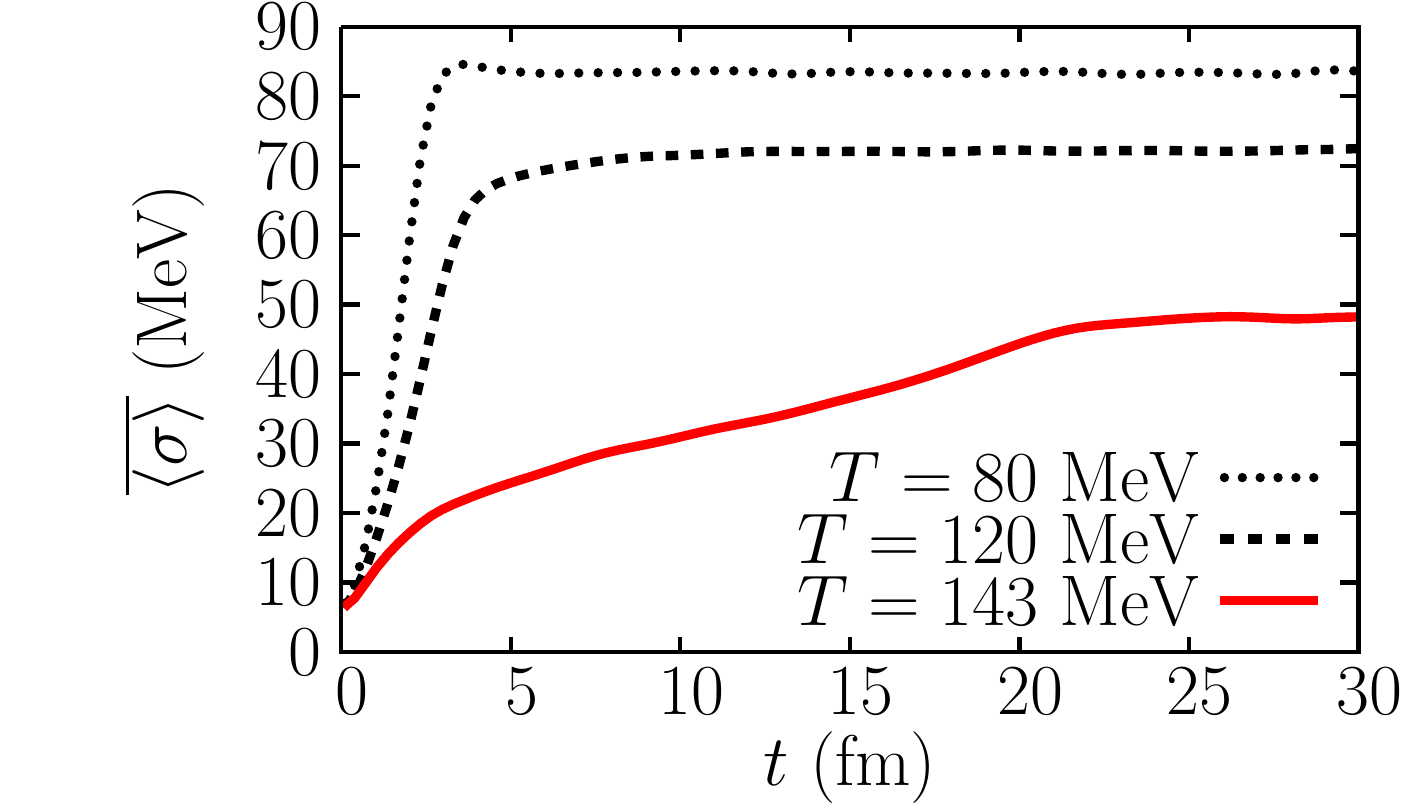}
  }
\quad
  \subfloat[\label{fig:cprelax}]{
  \includegraphics[scale=0.5]{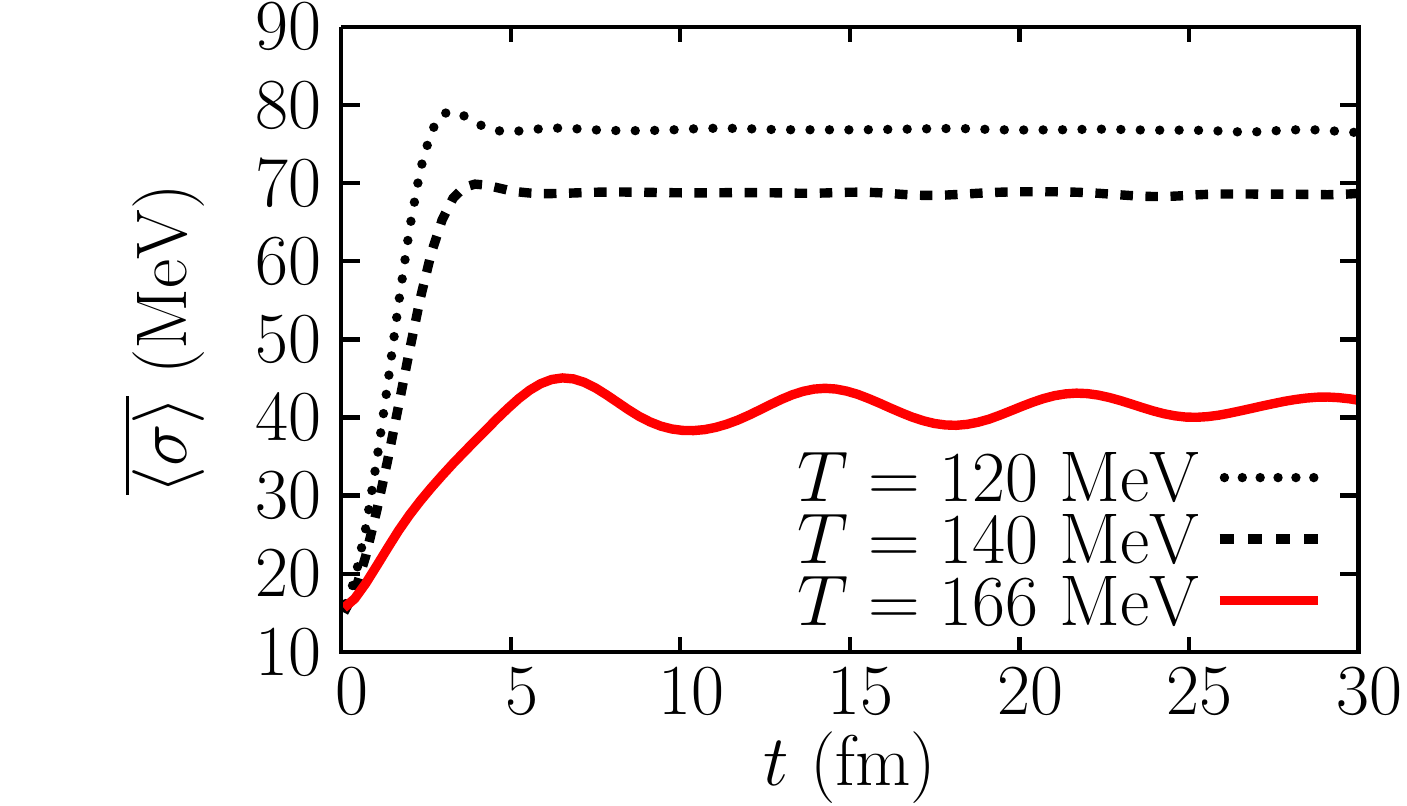}
  }
\caption{\ref{fig:forelax} Equilibration of the sigma field for several quench temperature $T<T_c$ through the first 
order transition. The barrier between the minima in the potential increases the relaxation time when the system 
relaxes near $T_c=172.9$~MeV. We chose $T_{\rm ini}=180$~MeV. \ref{fig:cprelax}  Equilibration of the sigma field for several quench 
temperature $T<T_c$ through the CP. Critical slowing down delays the dynamics and causes oscillations around the flat 
minimum when the system relaxes near $T_c=180.5$~MeV. We chose $T_{\rm ini}=186$~MeV.} 
\label{fig:sigmarelax}
\end{figure}

Results for both transition scenarios are shown in fig.\ \ref{fig:sigmarelax}. They show the evolution 
of the noise and volume averaged value of the sigma field in time for different quenching temperatures $T$. The solid curves indicate the case 
where the system relaxes near the corresponding transition temperature. In both cases the dynamics is slowed down, 
however for different reasons. At the first order phase transition the barrier that separates minima near $T_c$ 
is responsible for a significant delay in the relaxation dynamics. For the second order transition we observe critical 
slowing down. This is inherent in the model due to the vanishing of the damping coefficient around the critical 
temperature that causes the system to oscillate around the equilibrium state and prolong the relaxation time up to 
infinity.

\begin{figure}[htbp]
\centering
  \subfloat[\label{fig:loopforelax}]{
  \includegraphics[scale=0.5]{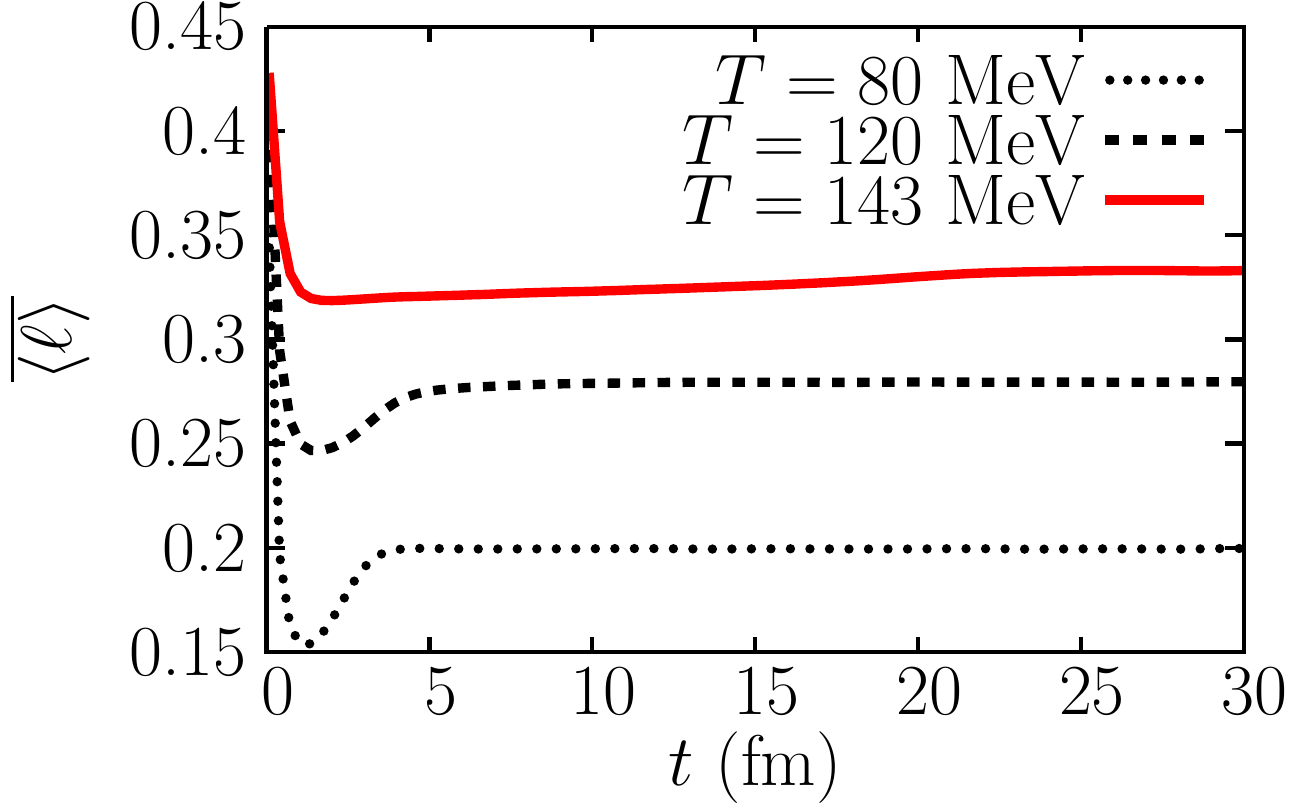}
  }
\qquad
  \subfloat[\label{fig:loopcprelax}]{
  \includegraphics[scale=0.5]{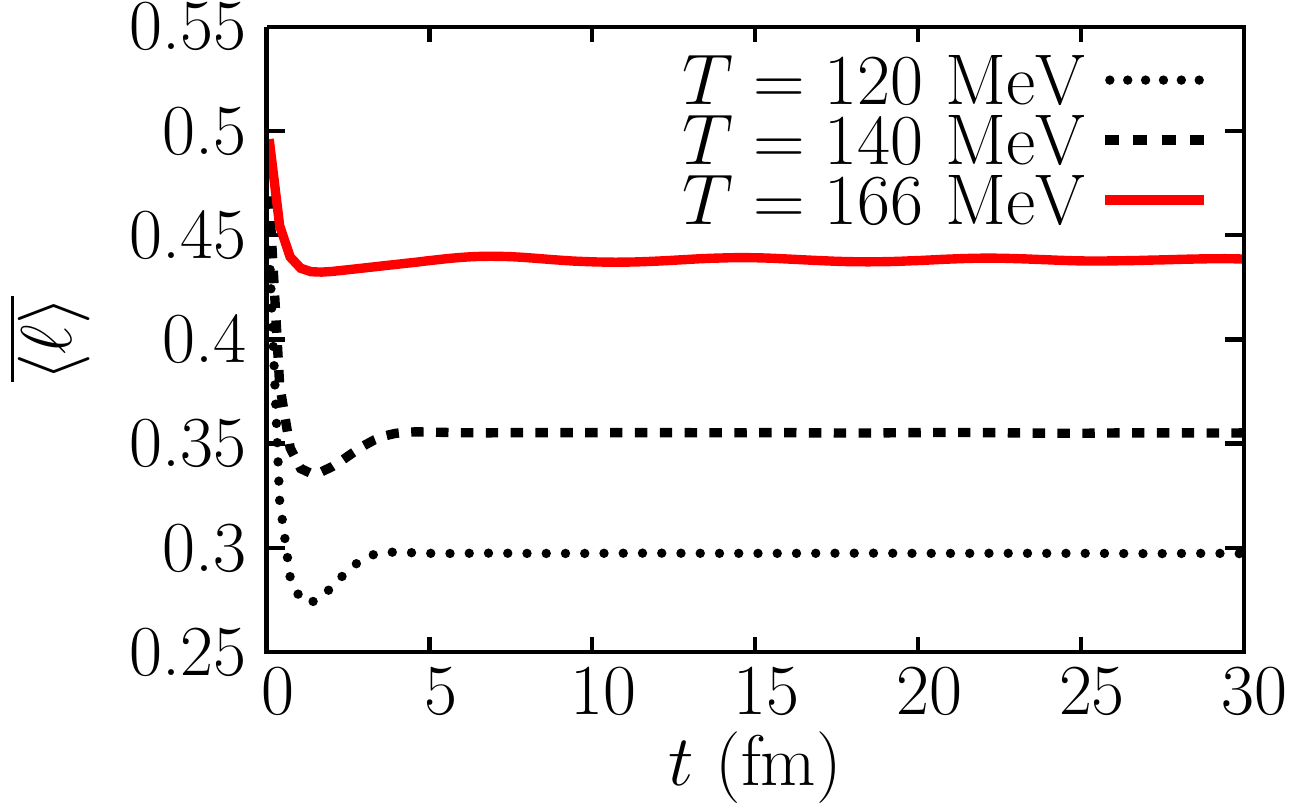}
  }
\caption{\ref{fig:loopforelax} Equilibration of the Polyakov loop for several temperature quenches $T<T_c$ through the 
first order transition. We chose $T_{\rm ini}=180$~MeV. \ref{fig:loopcprelax} Equilibration of the Polyakov loop for several 
temperature quenches $T<T_c$ through the CP. We chose $T_{\rm ini}=186$~MeV.}
\label{fig:looprelax}
\end{figure}

The average values of the Polyakov loop as a function of time are shown in fig.\ \ref{fig:looprelax}. 
Here we find again a prolongation of the relaxation time near the first order phase transition where the average value 
is slowly growing until $t=25$~fm while for the other quenching temperatures, the system is equilibrated 
after $5$~fm. For a CP scenario, we observe the same effect than for the sigma field: critical slowing down 
near the transition temperature, nevertheless with a small amplitude. 

We performed these simulations with various damping coefficients for the Polyakov loop $\eta_{\ell}$ ranging from 
$1/{\rm fm}$ up to $10/{\rm fm}$. A significant difference in the relaxational behavior could only be observed in the case where 
the system equilibrated near the first order phase transition where a larger value of $\eta_{\ell}$ caused a larger 
relaxation time. In all other cases the results are not sensitive to the choice of damping. Therefore we may consider 
our choice of $\eta_{\ell}=5/{\rm fm}$ as justified, especially for the case of an expanding hot medium, which we 
finally aim to describe. 

\begin{figure}[htbp]
\centering
  \subfloat[\label{fig:sig_t12}]{
  \includegraphics[scale=0.5]{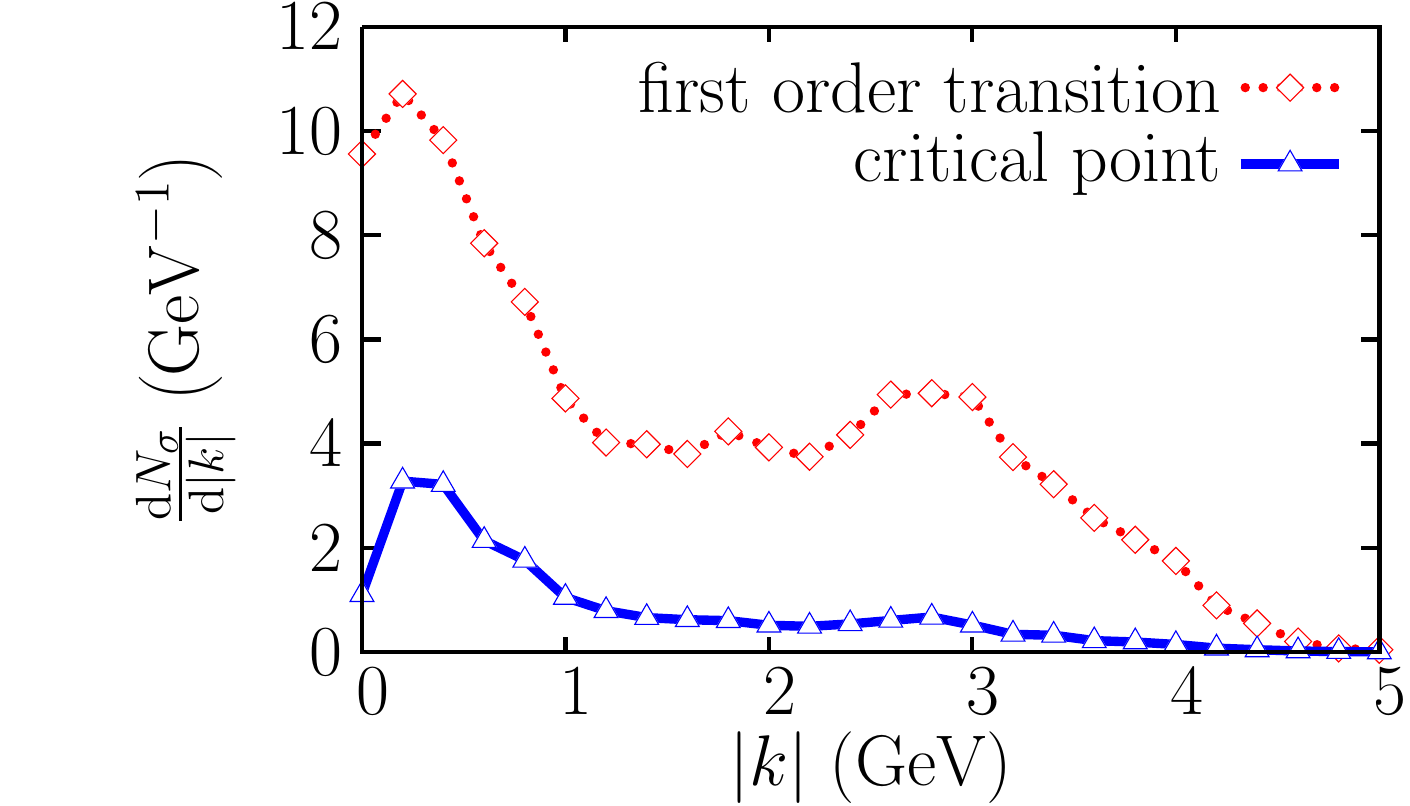}
  }
\qquad
  \subfloat[\label{fig:sig_t24}]{
  \includegraphics[scale=0.5]{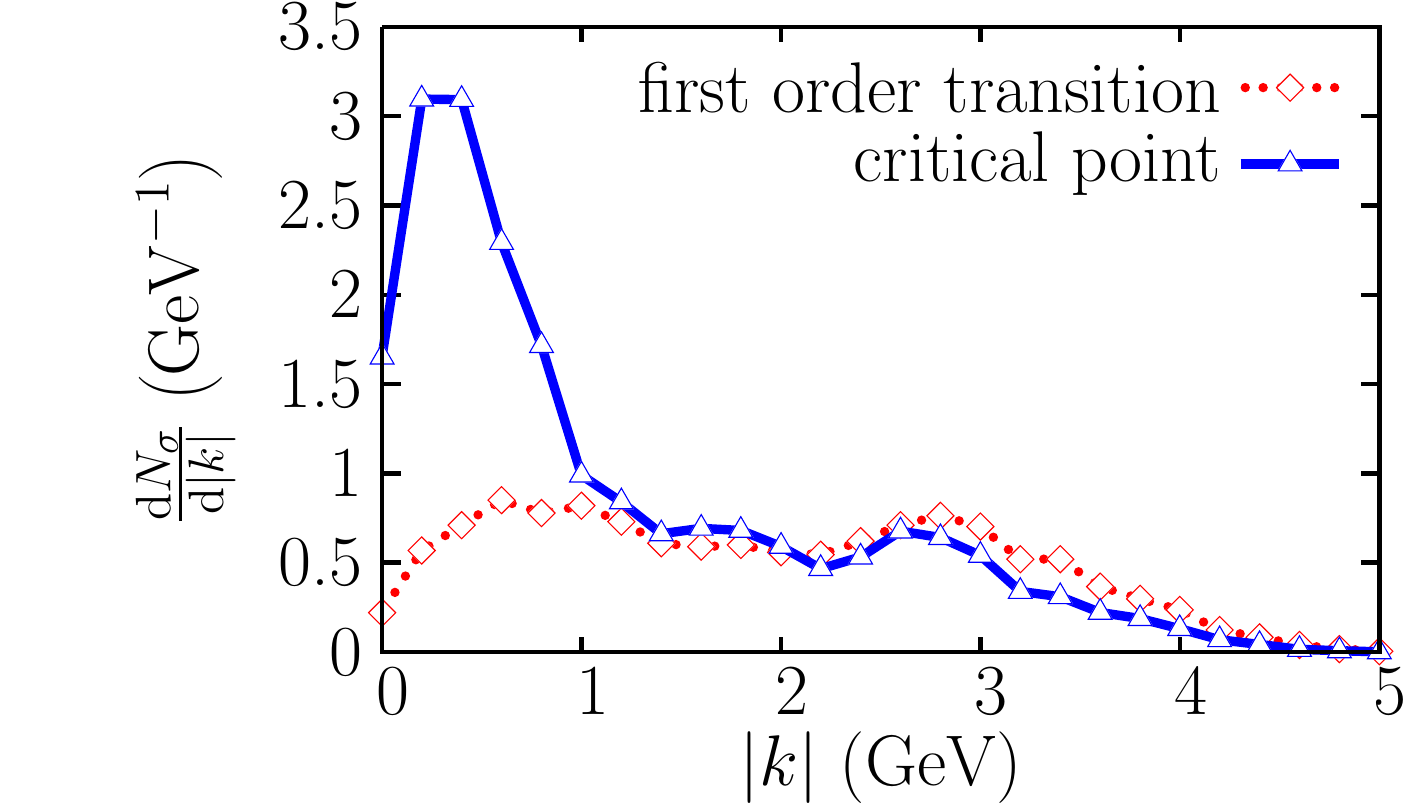}
  }

  \subfloat[\label{fig:lo_t12}]{
  \includegraphics[scale=0.5]{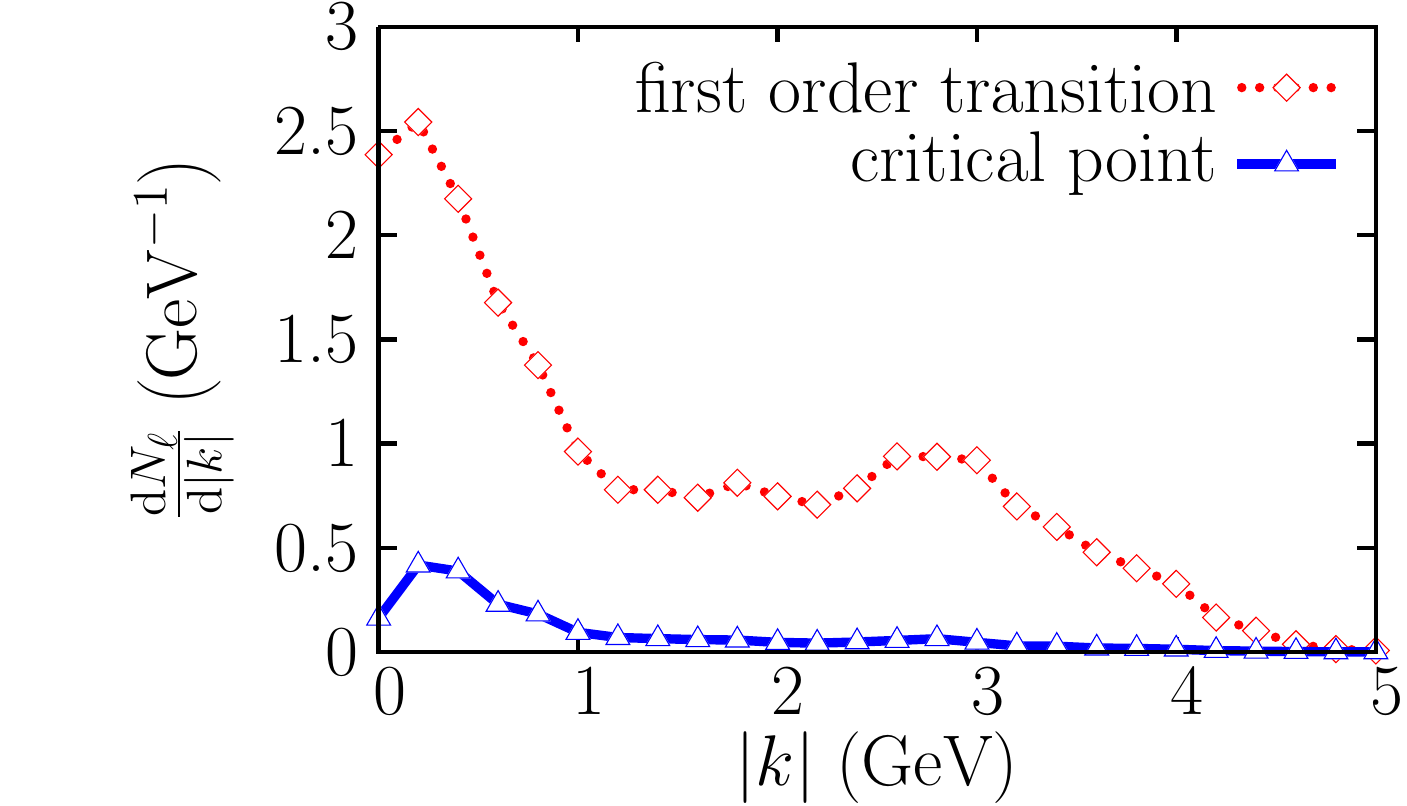}
  }
\qquad
  \subfloat[\label{fig:lo_t24}]{
  \includegraphics[scale=0.5]{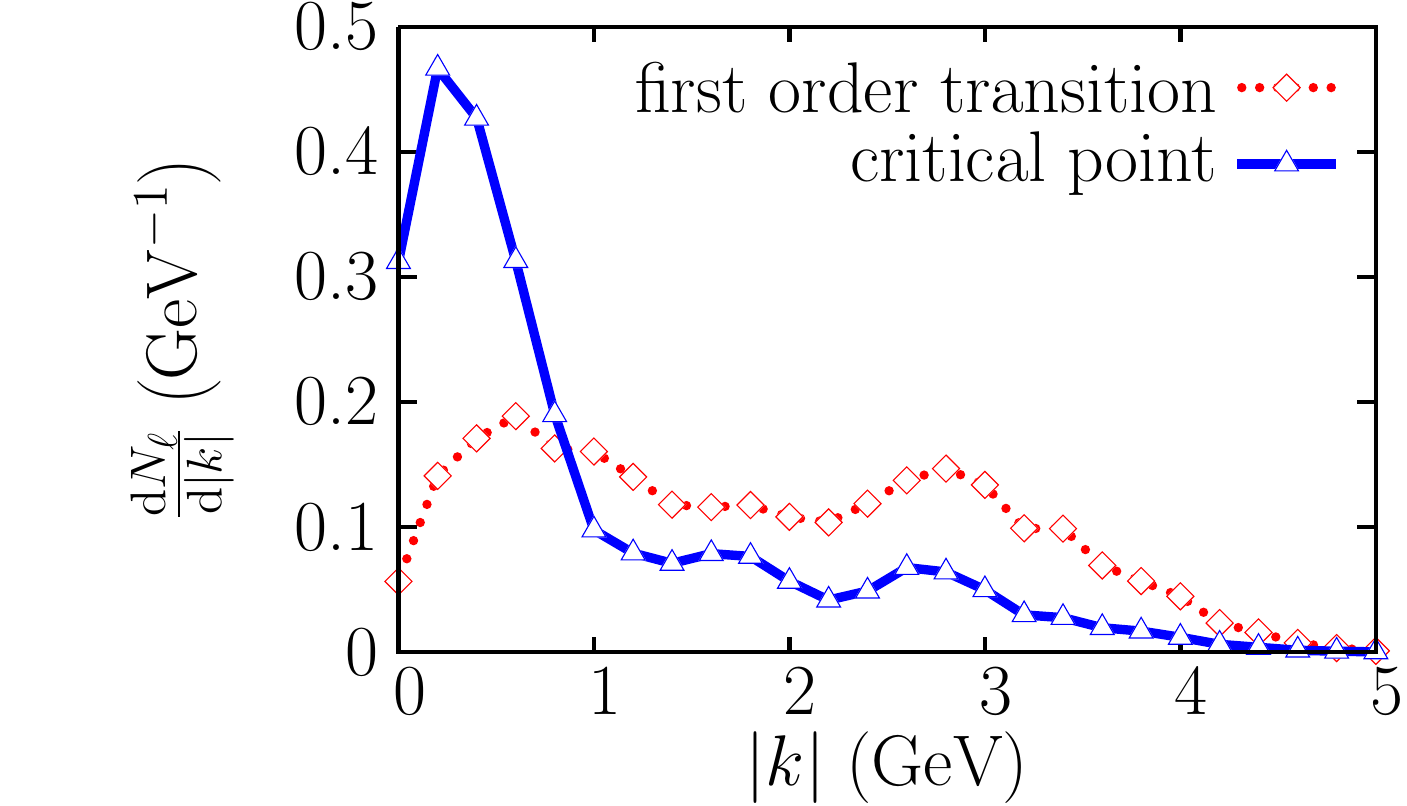}
  }
\caption{\ref{fig:sig_t12} Intensity of sigma fluctuations during the transition process at $t=12$~fm (first order) 
and $t=3$~fm (CP). \ref{fig:sig_t24} Intensity of sigma fluctuations after equilibration at $t=24$~fm. In the CP scenario we find 
an enhancement of the soft modes. \ref{fig:lo_t12} Intensity of Polyakov loop fluctuations during the transition process at 
$t=12$~fm (first order) and $t=3$~fm (CP). \ref{fig:lo_t24} Intensity of Polyakov loop fluctuations after equilibration at 
$t=24$~fm. In the CP scenario we find an enhancement of the soft modes.}
\label{fig:intensity}
\end{figure}

The fluctuations of the order parameters can be analyzed by calculating their intensity $N_{\sigma}$ and $N_{\ell}$.
For the sigma field this quantity is given by \cite{Abada:1996bw,AmelinoCamelia:1997in,arXiv:1105.1962}
\begin{equation}
\frac{\mathrm d N_{\sigma}}{\mathrm d^3 k}= \frac{a_k^{\dagger}a_k}{(2\pi)^3 2\omega_k}=\frac{\omega_k^2|\delta\sigma_k|^2+|\partial_t\sigma_k|^2}{(2\pi)^3 2\omega_k}~,
\end{equation}
where $a_k^{\dagger}$ and $a_k$ are the Fourier coefficients of the expansion of the sigma field around its equilibrium value
$\delta\sigma=\sigma-\sigma_{\rm eq}$ and of the conjugate momentum field $\partial_t \sigma$. The energy 
of the $k$-th mode is $\omega_k=\sqrt{\vec k^2+m_\sigma^2}$. 
We use an analogous definition for the Polyakov loop
\begin{equation}
\frac{\mathrm d N_{\ell}}{\mathrm d^3 k}=T^2 \frac{\omega_k^2|\delta\ell_k|^2+|\partial_t\ell_k|^2}{(2\pi)^3 2\omega_k}
\end{equation}
with $\omega_k=\sqrt{\vec k^2+m_\ell^2}$, although this field formally has no kinetic energy term 
(see discussion at the end of section \ref{sec:eqmotion}). 
In equilibrium the quantities $N_\sigma$, $N_\ell$ may be interpreted as particle numbers, but to avoid confusion we 
call them `intensities of fluctuations'.
Several histograms of $N_{\sigma}$ and $N_{\ell}$ as a function of the wave number $|k|$ evaluated at different times are shown 
in figs.\ \ref{fig:sig_t12}, \ref{fig:sig_t24} for the sigma 
field and figs.\ \ref{fig:lo_t12}, \ref{fig:lo_t24} for the Polyakov loop. The figures on the left hand side show the intensity 
at early times during the transition process. We see that here fluctuations at the first order transition are clearly 
enhanced compared to the CP scenario. On the right hand side the intensities are shown for the time 
$t=24$~fm after the system has 
equilibrated. Here we see that the long wavelength modes of both order parameters are enhanced at the CP, a 
typical and well-known critical phenomenon \cite{Stephanov:1999zu}.

\section{Dynamics in an expanding medium}
\label{sec:expandingmedium}

Here we are interested in the coupled dynamics of a system which is not confined in a box but freely expands into vacuum, 
similar to what happens after a heavy-ion collision. We study the relaxational behavior of the sigma field and Polyakov 
loop during the nonequilibrium evolution and investigate the influence of energy-momentum exchange on the temperature 
evolution in both transition scenarios.

\subsection{Numerical implementation}
For this simulation we use a $128^3$ grid and initialize in its center a droplet which is ellipsoidal in the x-y-plane 
and uniform in z-direction. This resembles the almond shape of the overlap region of two colliding nuclei. 
The droplet has a temperature of $T_{\rm ini}=200$~MeV, well above both transition temperatures and is 
smoothed by a Woods-Saxon distribution function at its edges:
\begin{equation}
 T(\vec x,t=0)=\frac{T_{\rm ini}}{\big( 1+\exp[(\tilde r-\tilde R)/\tilde a] \big)\big(1+\exp[(\left|z\right|-l_z)/\tilde a] \big)}~.
\end{equation}
Here, $\tilde r=\sqrt{x^2+y^2}$, $\tilde a=0.6$~fm is the thickness of the transition layer to vacuum and 
\begin{equation}
 \tilde R=  \begin{cases}
  \frac{a b \tilde r}{\sqrt{b^2x^2+a^2y^2}},  & \tilde r \neq 0\\
  a, & \tilde r = 0
  \end{cases}~.
\end{equation}
The ellipsoidal parameters are chosen as $a=r_A-\tilde b /2$ and $b=\sqrt{r_A^2-\tilde b/4}$, $r_A=6.5$~fm denoting 
the radius of the two nuclei and $\tilde b=6$~fm the impact parameter. 
The extent in z-direction is $2l_z=12$~fm. The sigma and Polyakov loop fields are initialized with their 
respective equilibrium distribution, then the energy density and pressure of the quark fluid are calculated.  
We choose an initial velocity profile with $v_z(\vec x,t)=|z|/l_z \cdot v_{\rm max}$ with $v_{\rm max}=0.2$. The transverse 
velocities $v_x$ and $v_y$ are set to zero in the beginning. 

\subsection{Energy conservation}
\label{sec:energy}
\begin{figure}[htbp]
\centering
  \subfloat[\label{fig:energyfo}]{
  \includegraphics[scale=0.5]{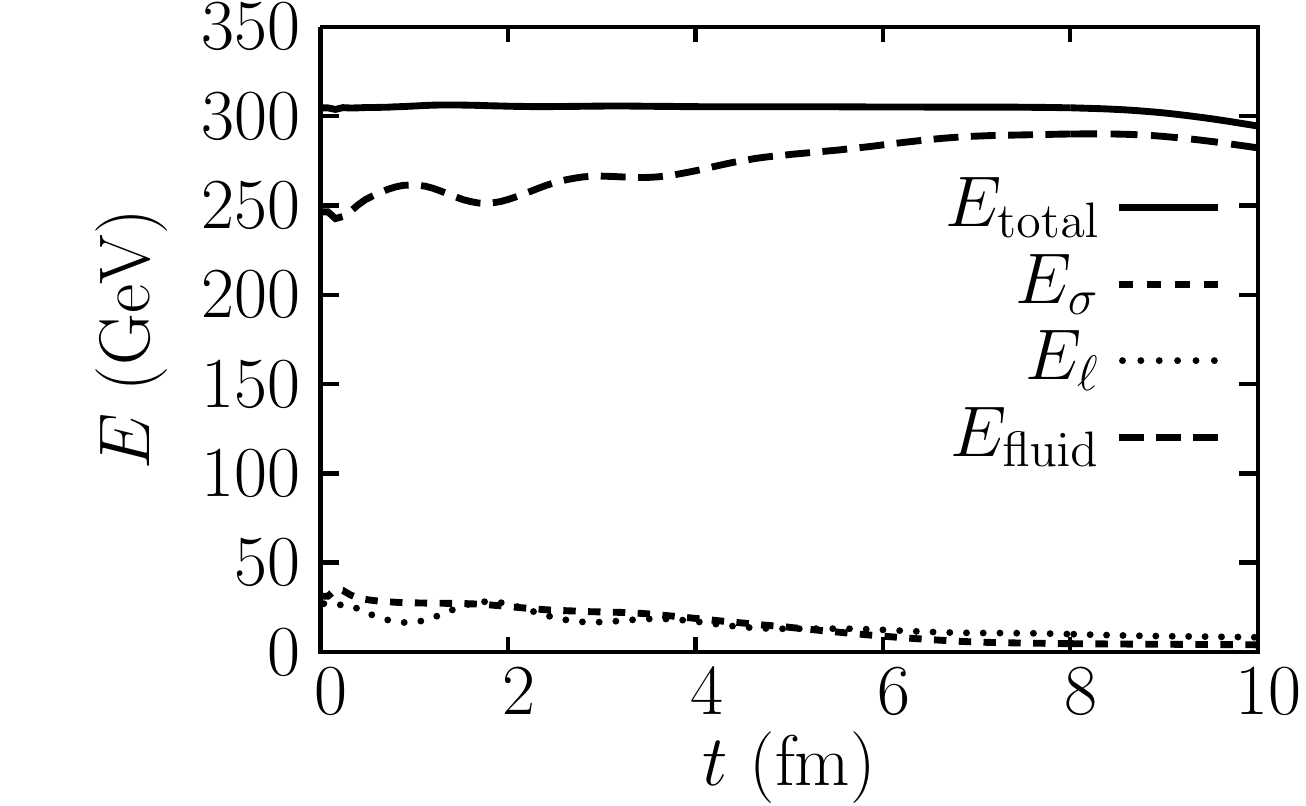}
  }
\qquad
  \subfloat[\label{fig:energycp}]{
  \includegraphics[scale=0.5]{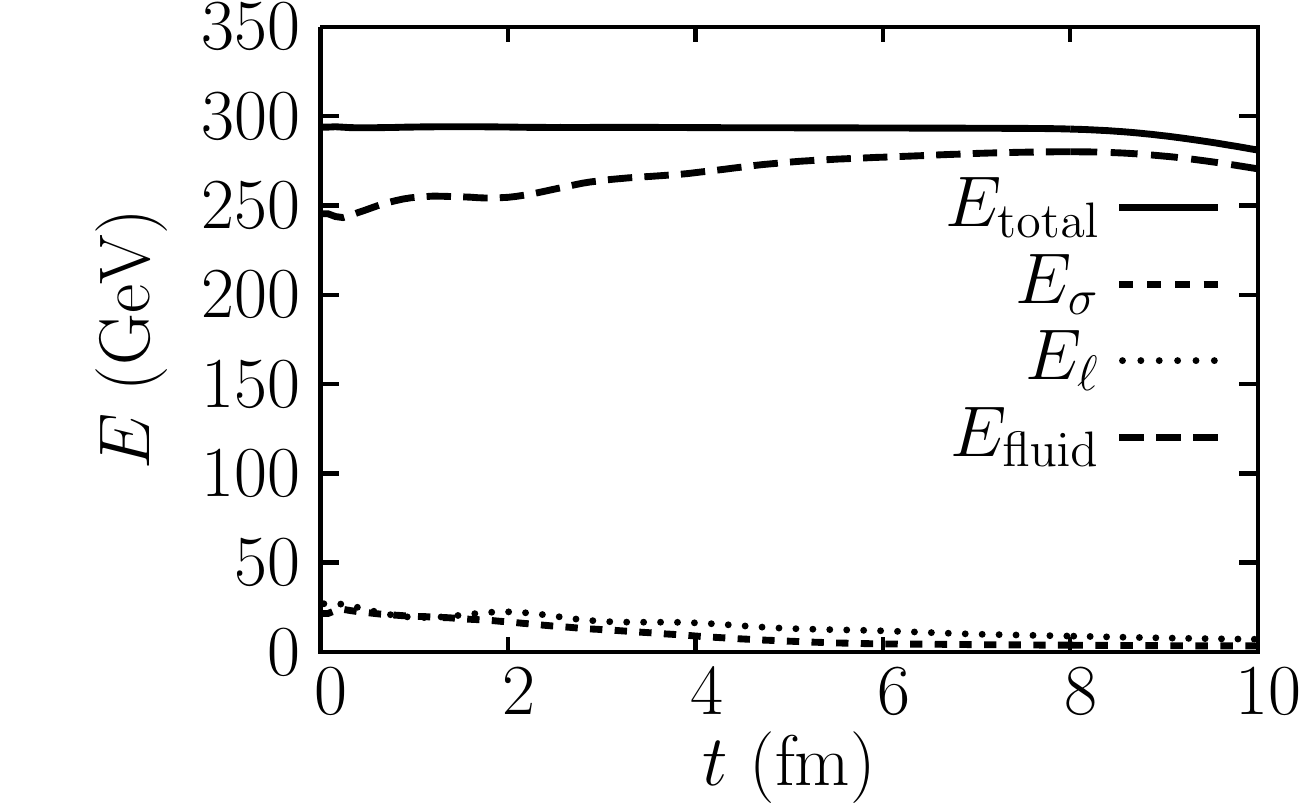}
  }
\caption{\ref{fig:energyfo} The total energy as composed of the energy of the sigma field, Polyakov loop and the quark fluid of an 
expanding system for a first order transition scenario. \ref{fig:energycp} The total energy as composed of the energy of the sigma 
field, Polyakov loop and the quark fluid of an expanding system for a scenario with a CP.}
\label{fig:energy}
\end{figure}

We let the system expand by full $3+1$ dimensional fluid dynamics
and check the conservation of the total energy throughout the evolution. The energy of the sigma and Polyakov loop 
field are given by (\ref{eq:energysigmafield}) and (\ref{eq:energypolyakovfield}).
Fig.\ \ref{fig:energy} shows the total energy and the partial energies of the quark fluid, the sigma field 
and the Polyakov loop during the fluid dynamical expansion. For both scenarios the total energy is well 
conserved until the quark fluid reaches the edges of the computational grid after $8$~fm. 

\subsection{Supercooling and reheating}
\begin{figure}[htbp]
\centering
  \subfloat[\label{fig:expfo}]{
  \includegraphics[scale=0.5]{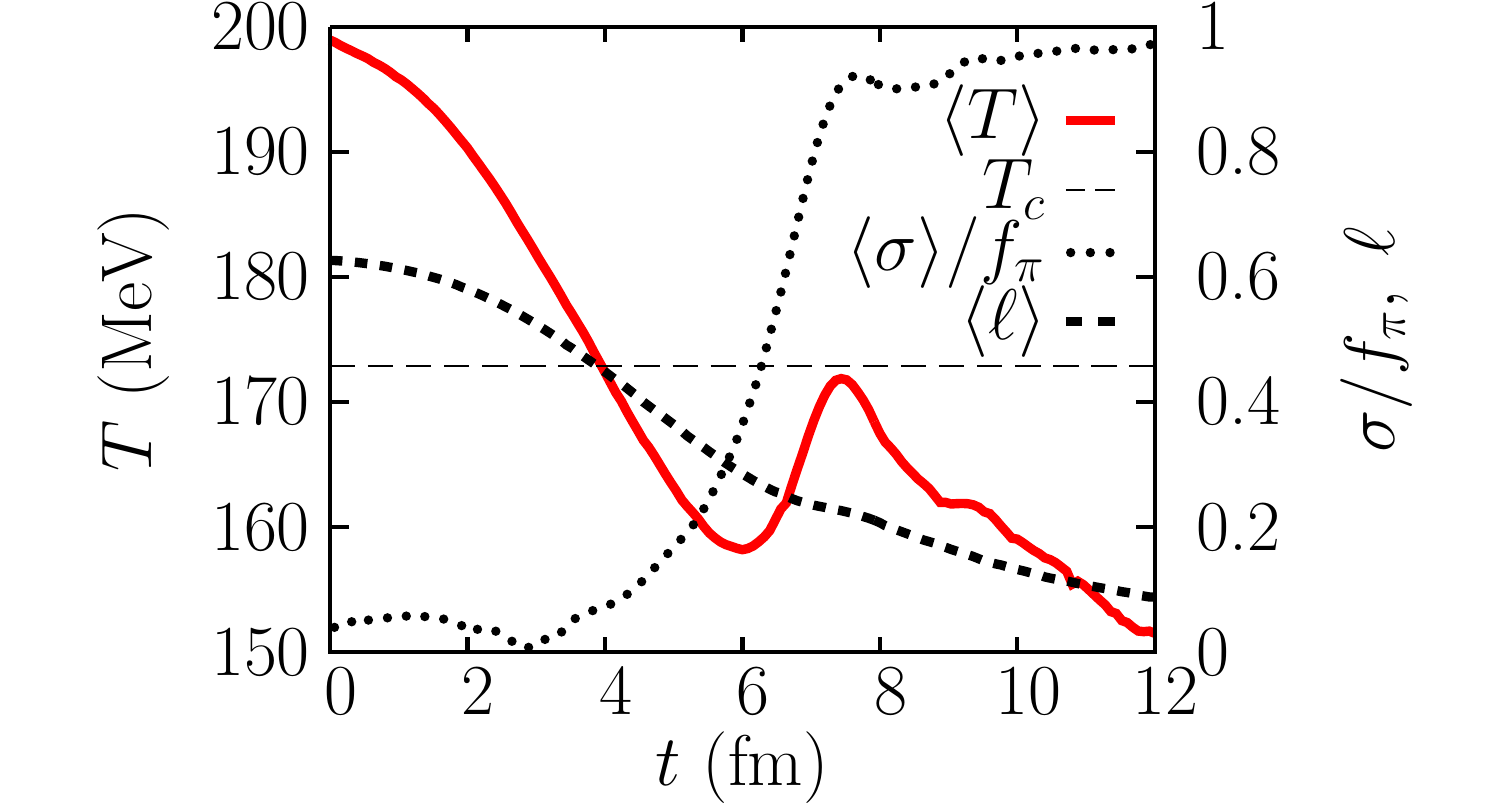}
  }
\quad
  \subfloat[\label{fig:expcp}]{
  \includegraphics[scale=0.5]{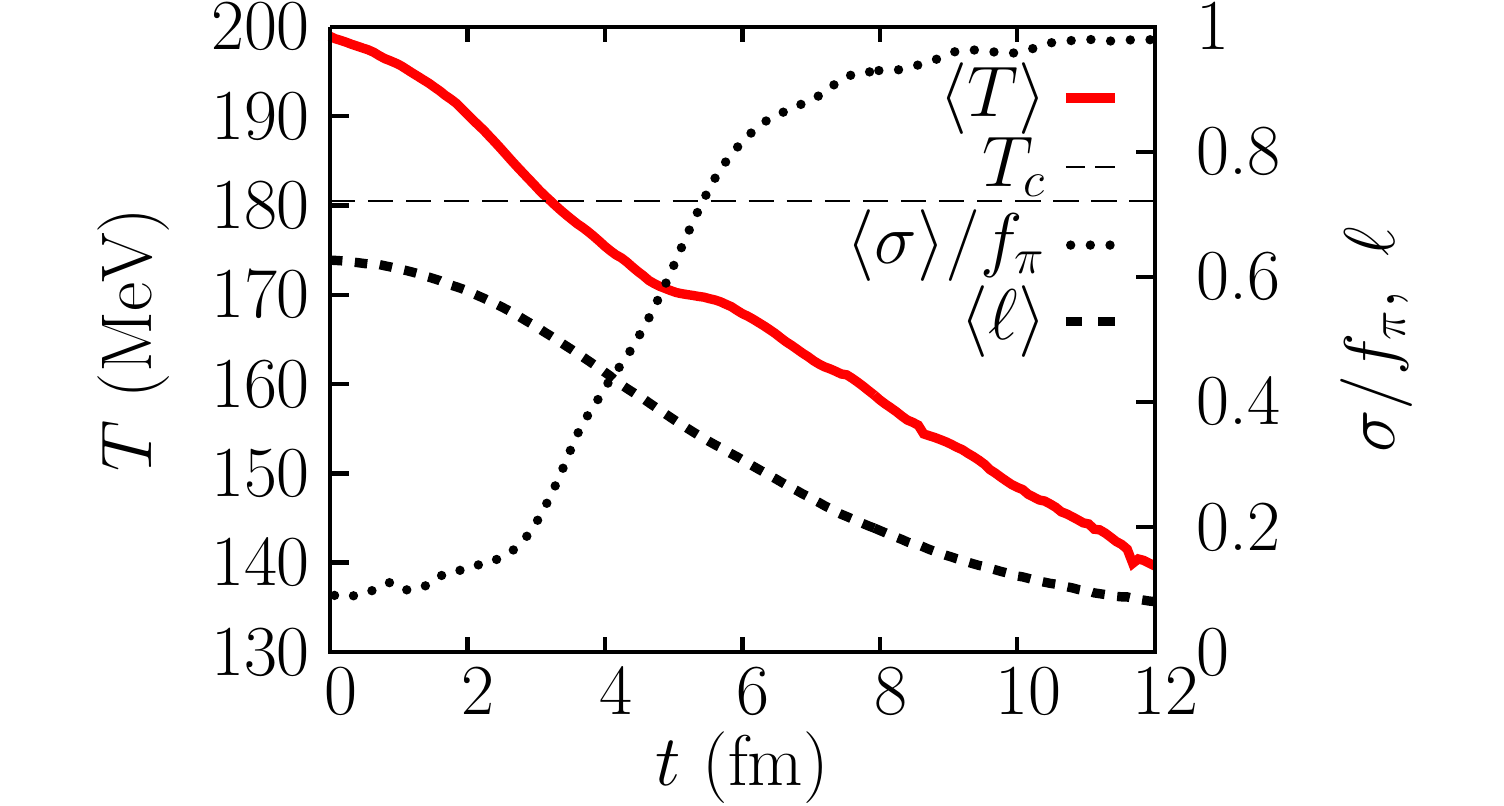}
  }
\caption{\ref{fig:expfo} Evolution of the average temperature, sigma field and Polyakov loop in a system evolving 
through the first order transition. Supercooling followed by reheating can be observed. The horizontal line denotes 
the critical temperature. \ref{fig:expcp} Evolution of the average temperature, sigma field and Polyakov loop in a system 
evolving through the CP. The temperature decreases monotonically with a small plateau slightly below $T_c$. The 
horizontal line denotes the critical temperature.}
\label{fig:exp}
\end{figure}

During the fluid dynamic expansion we extract the average temperature $\langle T\rangle$, sigma field $\langle\sigma\rangle$ 
and Polyakov loop $\langle\ell\rangle$ in a central cubic volume of $1\rm{fm}^3$ inside the 
hot matter as a function of time. The results are shown 
in fig.\ \ref{fig:expfo} for the first order transition and in fig.\ \ref{fig:expcp} for a scenario with a CP. 
One can observe significant differences in the evolution of the average temperatures between both scenarios: 
For the case of the first order transition, fig.\ \ref{fig:expfo}, a reheating occurs after $6$~fm as a consequence of the formation of a supercooled phase 
below the transition temperature. We see that as the average temperature falls below $T_c$, the average values of the
sigma field and Polyakov loop remain close to their high temperature values around $\sigma/f_{\pi}=0.1$ and $\ell=0.4$. 
This supercooled state decays after about $2$~fm to the global minimum and transfers its energy into the fluid which 
consequently causes an increase in the average temperature.

For the CP scenario, fig.\ \ref{fig:expcp}, no reheating effect is observed. The temperature decreases monotonically with only a small 
plateau well below the transition temperature where the dynamics slightly slows down due to the flat shape of the 
effective potential. The evolution of the averaged fields, especially for $\langle\sigma\rangle$, proceeds less rapidly 
than at the first order phase transition. 

\subsection{Domain formation}

\begin{figure}[htb!]
\centering
\includegraphics[scale=0.55]{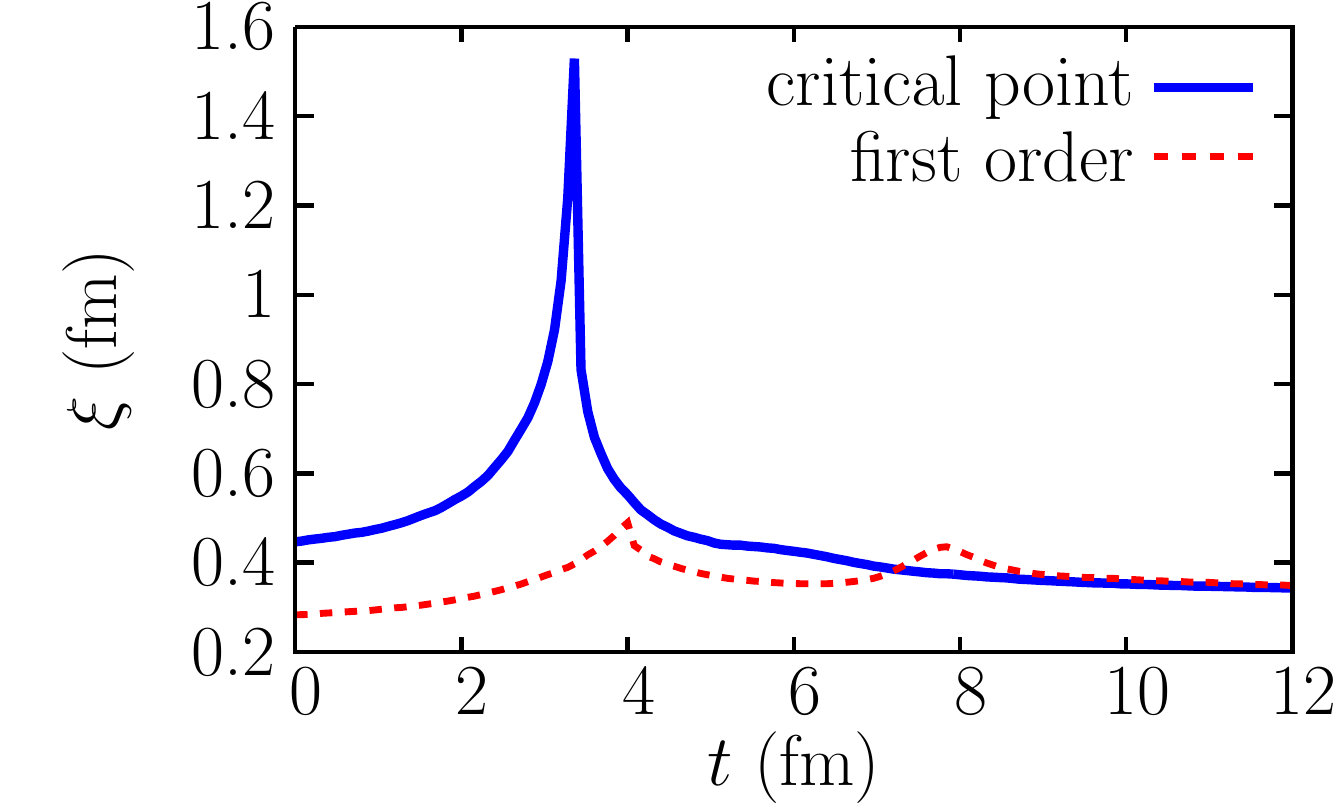}
\caption{ Correlation length of the sigma field as a function of time, for a CP and a first order transition scenario.}
\label{fig:corrlength}
\end{figure}

In our previous calculations we used the correlation functions (\ref{eq:correlationnoise}) and (\ref{eq:dissfluctpolyakov}) 
assuming that the stochastic noise fields in neighboring cells are not correlated as expressed by the spatial delta 
function. This leads to results that correspond to averaging over many events, see \cite{arXiv:1105.1962}. However, 
to better understand the differences between the two scenarios it is instructive to compare the microscopic structure 
of individual events. This requires a more consistent treatment of the field fluctuations by implementing realistic 
correlation lengths. 
Fig.\ \ref{fig:corrlength} shows the correlation length of the sigma field $\xi=1/m_{\sigma}$ as a function of time for the expansion 
through both the CP and the first order phase transition. Here, $m_{\sigma}$ is calculated out of the volume averaged temperature 
in the previous section. The effect of such an averaging on the correlation length in inhomogeneous 
systems has been discussed in \cite{Nahrgang:2012qm}.
For the first order transition, $\xi$ lies in a range of $0.3-0.5$~fm for the 
whole evolution while in the CP scenario it reaches a peak of about $1.5$~fm when the system crosses the transition 
temperature after $t=3.2$~fm, cf. fig.\ \ref{fig:expcp}. 
 
Below we present results obtained by correlating the noise fields over the spacelike distance $\xi=1/m_{\sigma}$ in each 
time step. The numerical procedure that implements such correlations has been described in \cite{Paech:2003fe}. 
For each spatial cell we perform an averaging of the randomly distributed noise field over a surrounding cube of linear size 
$\xi=n\Delta x$ via
\begin{equation}
 \xi_{\sigma}'(\vec x)=\frac{1}{n^3}\sum_{i,j,k=0,...,n-1}\xi_{\sigma}(\vec x+i\Delta x\vec e_1+j\Delta x\vec e_2+k\Delta x\vec e_3)~.
\end{equation}
As this procedure weakens the fluctuations, i.~e. $\langle\delta\xi_{\sigma}'^2\rangle\neq\langle\delta\xi_{\sigma}^2\rangle$, 
we have to rescale the noise field in order to obtain again the initial distribution width:
\begin{equation}
 \xi_{\sigma}''(\vec x)=\xi_{\sigma}'(\vec x)\sqrt{\frac{\langle\delta\xi_{\sigma}^2\rangle}{\langle\delta\xi_{\sigma}'^2\rangle}}~.
\end{equation}

\begin{figure}[htbp]
\centering
  \subfloat[\label{fig:sigma_fo_t4}]{
  \includegraphics[scale=0.6]{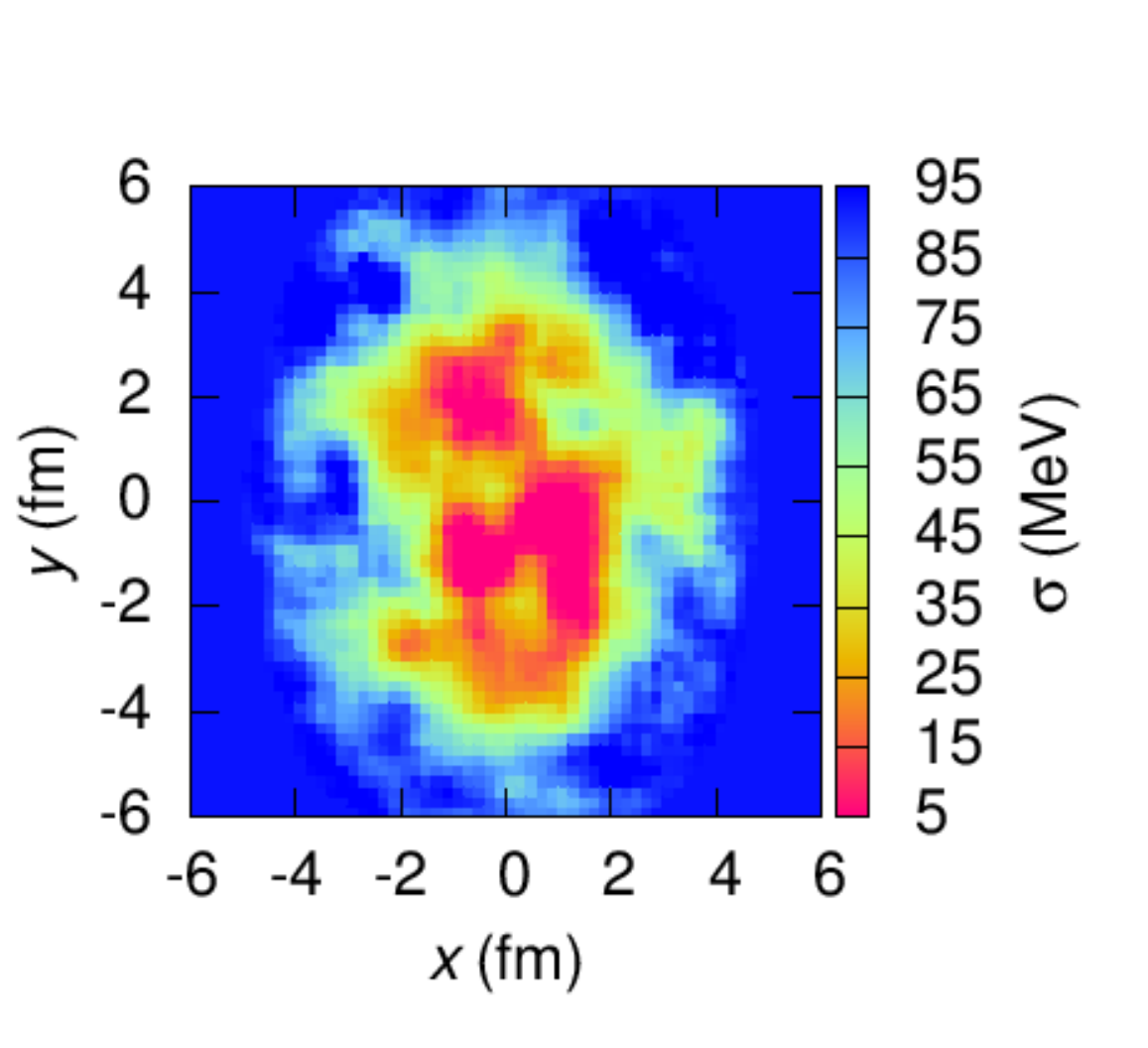}
  }
\qquad
  \subfloat[\label{fig:sigma_cp_t32}]{
  \includegraphics[scale=0.6]{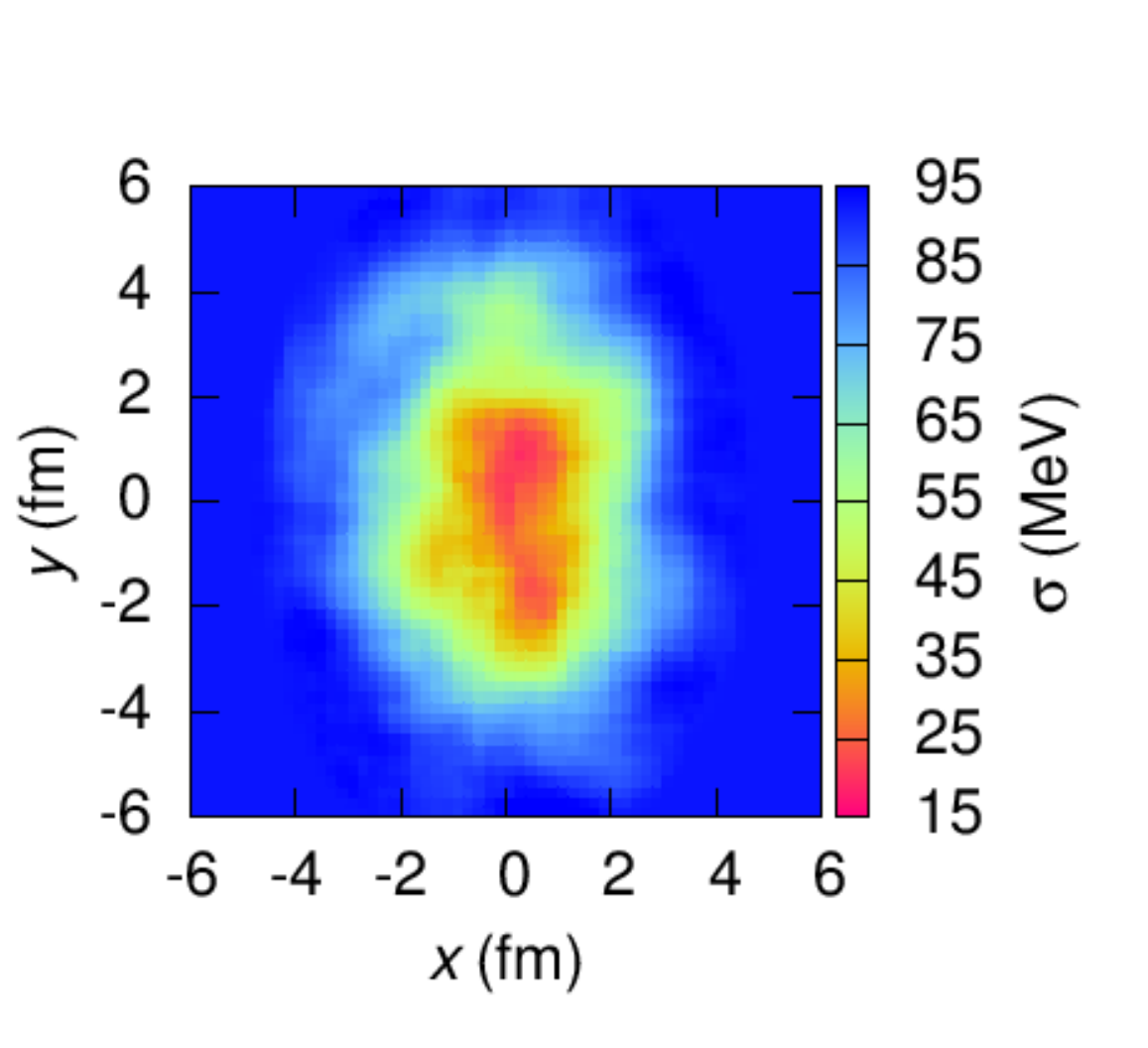}
  }
\caption{\ref{fig:sigma_fo_t4} Sigma field for $z=0$ at $t=4$~fm in a first order phase transition scenario.
\ref{fig:sigma_cp_t32} Sigma field for $z=0$ at $t=3.2$~fm in a scenario with a CP.}
\label{fig:sigma_focp}
\end{figure}

\begin{figure}[htbp]
\centering
  \subfloat[\label{fig:loop_fo_t4}]{
  \includegraphics[scale=0.6]{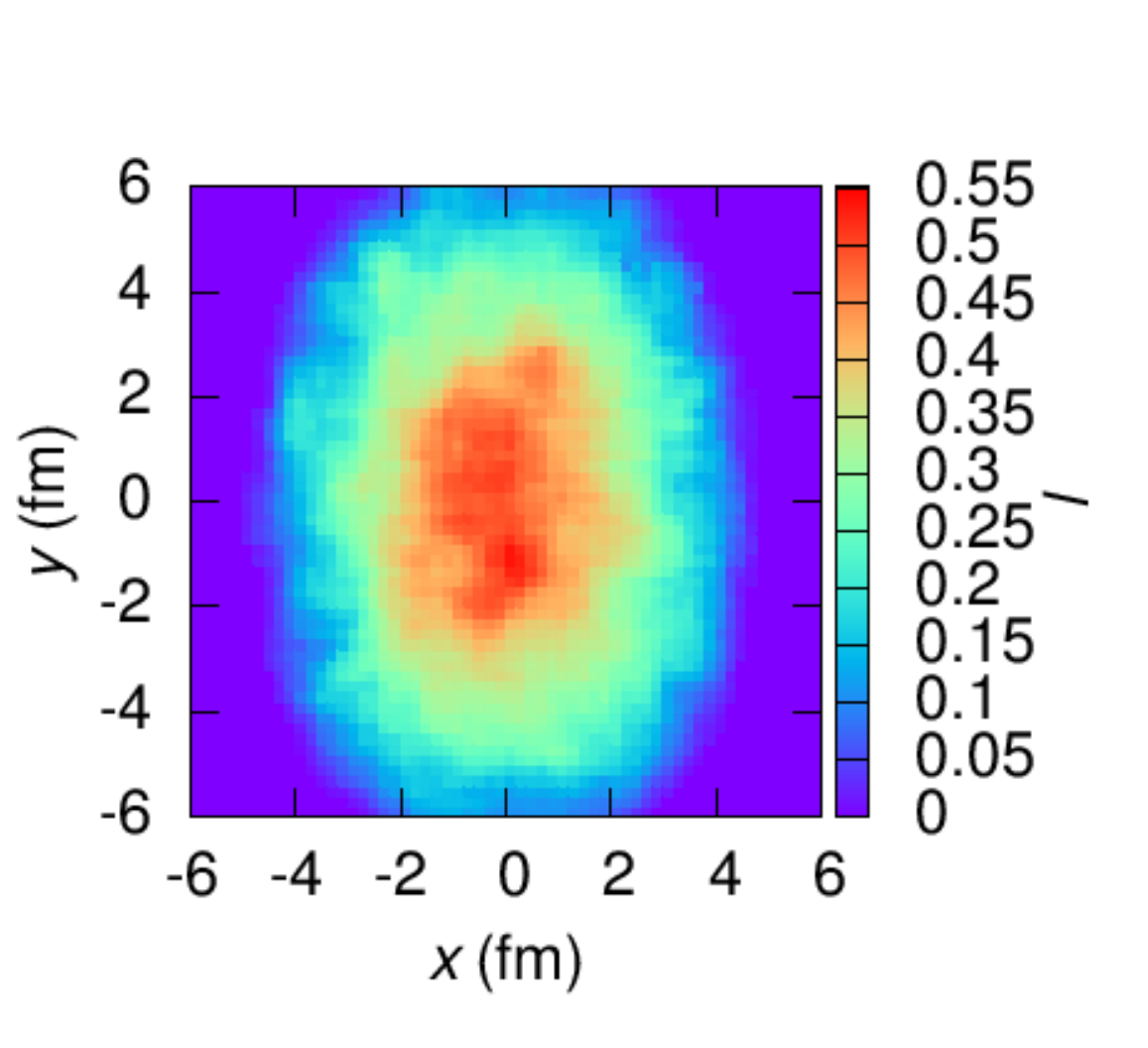}
  }
\qquad
  \subfloat[\label{fig:loop_cp_t32}]{
  \includegraphics[scale=0.6]{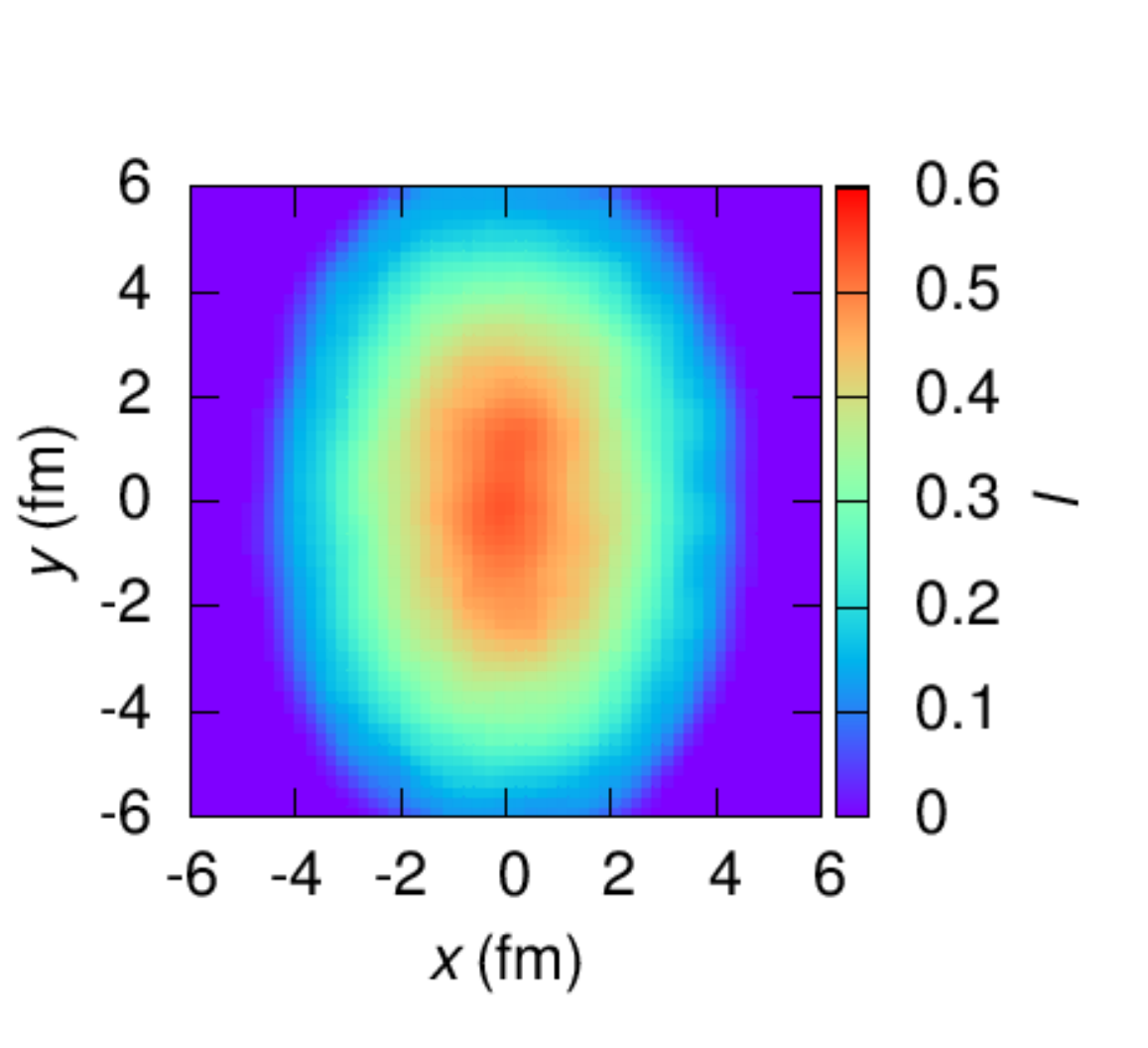}
  }
\caption{\ref{fig:loop_fo_t4} Polyakov loop for $z=0$ at $t=4$~fm in a first order phase transition scenario. 
\ref{fig:loop_cp_t32} Polyakov loop for $z=0$ at $t=3.2$~fm in a scenario with a CP.}
\label{fig:loop_focp}
\end{figure}

\begin{figure}[htbp]
\centering
  \subfloat[\label{fig:e_fo_t4}]{
  \includegraphics[scale=0.6]{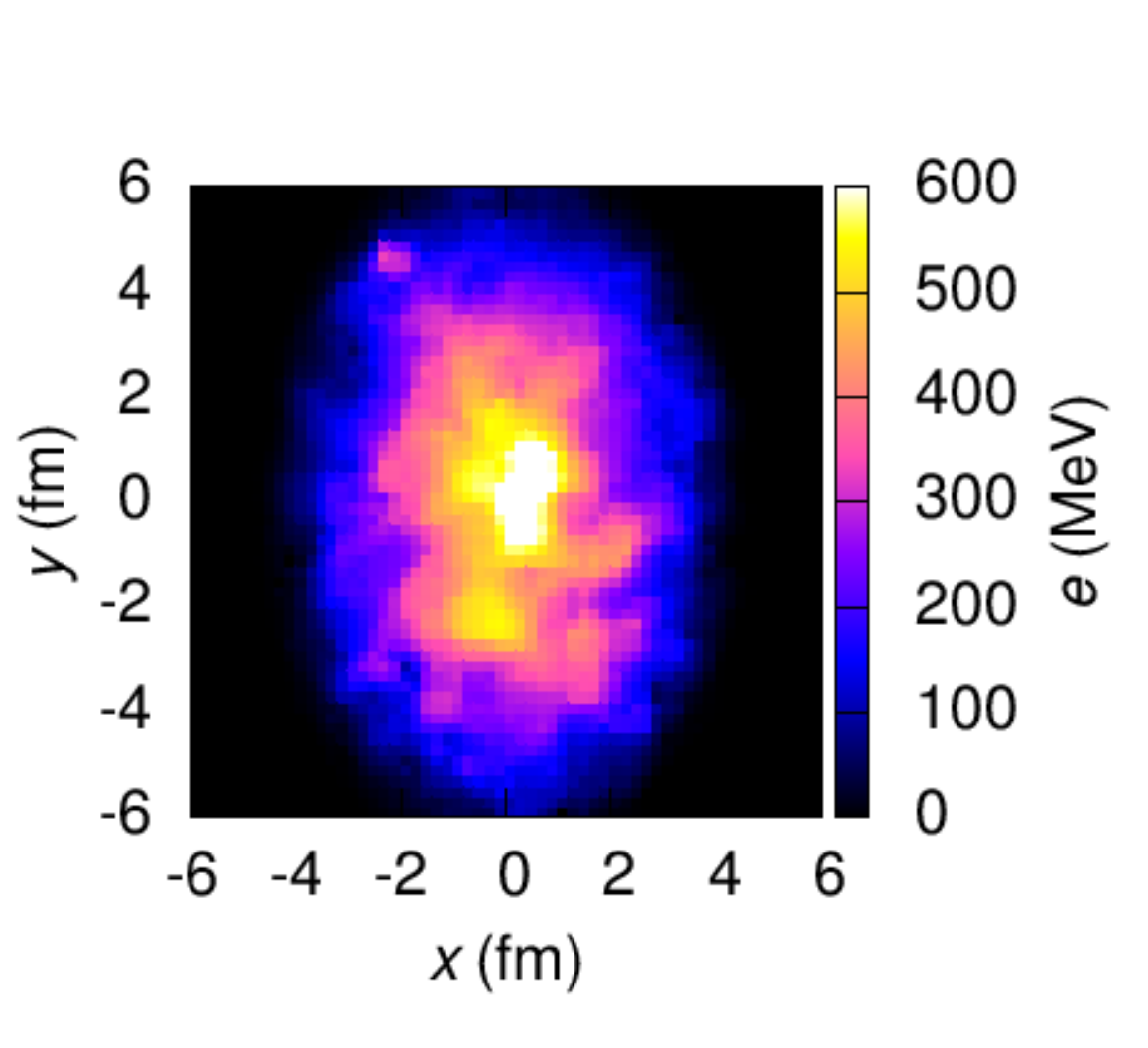}
  }
\qquad
  \subfloat[\label{fig:e_cp_t32}]{
  \includegraphics[scale=0.6]{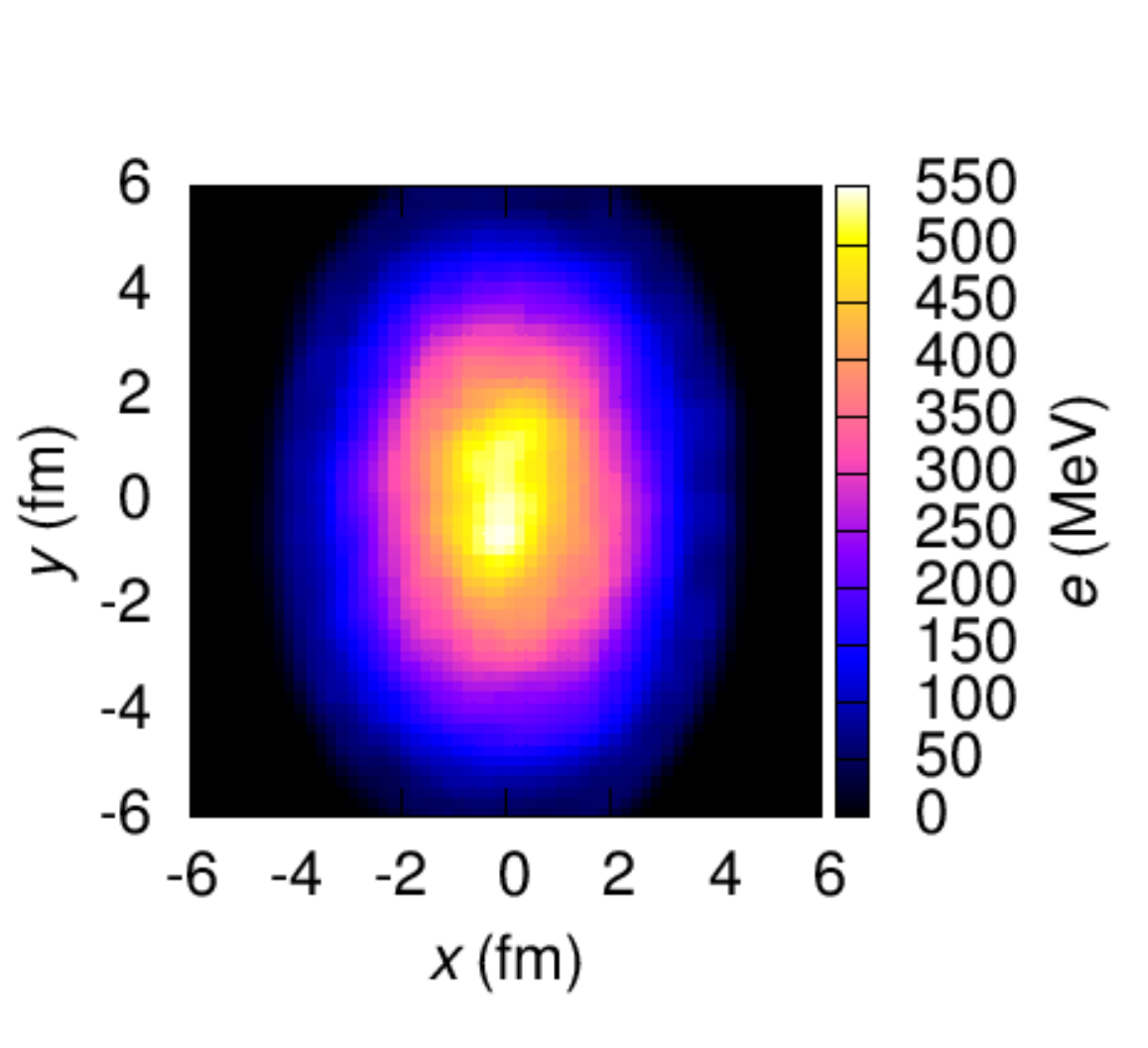}
  }
\caption{\ref{fig:e_fo_t4} Energy density for $z=0$ at $t=4$~fm in a first order phase transition scenario. 
\ref{fig:e_cp_t32} Energy density for $z=0$ at $t=3.2$~fm in a scenario with a CP.}
\label{fig:e_focp}
\end{figure}

An analogous correlation procedure is done for the Polyakov loop, correlating $\xi_{\ell}$ over the distance $1/m_{\ell}$. 
The following figures show spatial distributions in the transverse plane ($z=0$),  
for the sigma field in fig.\ \ref{fig:sigma_focp}, the Polyakov loop in fig.\ \ref{fig:loop_focp} and for the energy 
density in fig.\ \ref{fig:e_focp}. 
We chose a time of $t=4$~fm for the first order scenario corresponding to the onset of the transition 
process where $\langle T\rangle=T_c$, cf. fig.\ \ref{fig:expfo}. Here we expect to observe phase coexistence, i.~e. 
domains of the chirally broken phase in the chirally symmetric background or vice versa. For the CP scenario we chose a time of $t=3.2$~fm, where again 
$\langle T\rangle=T_c$, cf. fig.\ \ref{fig:expcp}, and furthermore the correlation length reaches its maximum value 
as shown in fig.\ \ref{fig:corrlength}. In this scenario there are no degenerate or metastable phases around $T_c$ but 
one single equilibrium state for each temperature. We therefore expect a more regular structure with no domains.

By inspecting the figures we indeed find striking 
difference between the two transition scenarios. At the first order transition, both the 
fields and fluid evolve irregularly. This effect is best observed in the 
sigma field, fig.\ \ref{fig:sigma_fo_t4}, where one can see the expected domains of the chirally broken phase embedded in the 
chirally symmetric background, see for instance the region around $(x,y)=(2,2)$. Here, the system produces large 
fluctuations in small spatial regions that are able to overcome the potential barrier and create these characteristic structures. 
Nevertheless, the different phases are connected smoothly, there are no sharp boundaries between the domains. 
This is a result of the Laplacian in the equation of motion which smoothens the gradients. 
Also in the Polyakov loop, fig.\ \ref{fig:loop_fo_t4}, and even more
in the energy density, fig.\ \ref{fig:e_fo_t4}, we observe a bumpy and irregular structure. Regions of high energy density 
are embedded in a background of lower energy density in the periphery, like for instance around $(x,y)=(-2,5)$.
These embedded regions are then going to 
grow and merge, leaving small islands of the symmetric phase which gradually shrink until the whole system is in the low 
temperature chirally broken phase. 

On the other hand, the evolution through the CP proceeds smoothly, the regular ellipsoidal structure is preserved. Especially 
at the transition point, where the correlation length grows large, we see a smooth shape in both the fields and the 
energy density of the quark fluid, see figs.\ \ref{fig:sigma_cp_t32}, \ref{fig:loop_cp_t32} and \ref{fig:e_cp_t32}. 
This striking difference in the event structure should manifest itself in the experimental data, e.~g. in the 
non-statistical multiplicity fluctuations of produced hadrons \cite{Misshustin:2007jx}.

\section{Summary and Outlook}
\label{sec:summary}
We presented a fully dynamical model to study the chiral and deconfinement phase transitions of QCD simultaneously. 
In our previous studies of chiral fluid dynamics we have derived the coupled dynamics of the sigma field and 
the quark heat bath within a self-consistent 2PI effective action approach. The results were now extended for the Polyakov 
loop so that both order parameters are propagated via Langevin equations taking into account the interaction with the 
quark heat bath via dissipation and noise. We assume that the structure of the source term for the Polyakov loop is 
analogous to that for the sigma field. During all simulations the total energy is well conserved.

We studied the relaxational behavior of the coupled system for different quench scenarios in a box. 
For both the first order and the CP scenario, relaxation near the transition point is delayed. 
At the first order phase transition the relaxation process is significantly delayed due to the barrier in the thermodynamic 
potential. The transition actually starts only when this barrier disappears. Near the CP, when the sigma mass drops to zero and therefore the damping 
vanishes, we observe perpetual oscillations of the sigma field around the equilibrium value. These fluctuations are 
also visible in the Polyakov loop, although with a small amplitude. Performing a Fourier analysis of the fluctuations for 
both order parameter fields we have observed another interesting peculiarity. While during the transition process the 
intensity of fluctuations is much stronger in a first order than in a CP scenario, the soft modes are stronger 
enhanced near the CP as compared with the first order transition if the systems are allowed to equilibrate.

For the evolution of an expanding fluid through the first order transition we have found a clear evidence for the formation of a 
supercooled phase. Its decay later on leads to a substantial reheating of the quark fluid. That is in contrast to the 
simulation with a CP where the temperature decreases monotonically. Furthermore, during the onset of the first order 
transition, small 
domains of different phases coexist and create inhomogeneities in the energy density. In the CP scenario, where the
correlation length grows large near the critical temperature, we observe a more homogeneous structure in the fields 
and fluid. A detailed analysis of domain formation at the first order phase transition will be provided in future work. 
Furthermore, we will extend this model to finite baryochemical potential and study trajectories in the full 
$T$-$\mu$-plane as well as fluctuations of baryon number densities. 

We think about several extensions and refinements of the model. The damping coefficient for the sigma field 
should include not only the $\sigma\leftrightarrow q\bar q$ decay but also the decay into pions. The soft modes 
which are linked to the order parameter of the chiral transition are furthermore affected by interaction with the hard 
modes which would give another contribution to the heat bath and the dissipation process. For the Polyakov loop we used a purely 
phenomenological Langevin equation with a constant damping coefficient. This needs improvement, for instance by extracting 
this term from Monte Carlo simulations like it has been proposed in \cite{Fraga:2007gg}. The present studies will allow for 
a better quantitative understanding of the signatures of the CP especially at FAIR energies.

\section*{Acknowledgements}
The authors thank Stefan Leupold and Carsten Greiner for fruitful discussions and Dirk Rischke for providing the SHASTA code. 
This work was supported by the Hessian LOEWE initiative Helmholtz International Center for FAIR. I.~M.
acknowledges partial support from the grant NSM-215.2012.2 (Russia).

\end{document}